\documentclass{article}

\usepackage{PRIMEarxiv}
\usepackage[utf8]{inputenc} 
\usepackage[T1]{fontenc}    
\usepackage{hyperref}       
\usepackage{url}  
\usepackage{booktabs}       
\usepackage{amsfonts}       
\usepackage{nicefrac}       
\usepackage{microtype}      
\usepackage{lipsum}
\usepackage{fancyhdr}       
\usepackage{graphicx}       
\usepackage{xcolor}
\usepackage{graphicx}
\usepackage{graphics}
\usepackage{subcaption}
\usepackage{amsmath}
\usepackage[version=4]{mhchem}
\usepackage{siunitx}
\usepackage{longtable,tabularx}
\usepackage{algorithm}
\usepackage{algpseudocode}
\usepackage{amsmath}
\usepackage{amssymb}
\usepackage{proof}
\usepackage{epsfig}
\usepackage{calc}
\usepackage{smartdiagram}
  \usesmartdiagramlibrary{additions} 
\graphicspath{{media/}}     
\newtheorem{problem}{Problem}

\newtheorem{definition}{\textnormal{\textbf{Definition}}}
\newtheorem{remark}{\textnormal{\textbf{Remark}}}
\newtheorem{theorem}{\textnormal{\textbf{Theorem}}}

\usepackage{tikz}
\usetikzlibrary{arrows,quotes} 
\RenewDocumentCommand{\smartdiagramconnect}{m m}{%
    \begin{tikzpicture}[remember picture,overlay]
    \foreach \start/\end in {#2}
    \draw
    (\start) edge[ #1] (\end);
    \end{tikzpicture}
}
        
\title{{Neural Koopman Lyapunov Control}
\thanks{Vrushabh Zinage (graduate student) and Efstathios Bakolas (Associate Professor) are with the Department of Aerospace Engineering and Engineering Mechanics,
The University of Texas at Austin, Austin, Texas 78712-1221, USA, {Emails: \tt\small vrushabh.zinage@utexas.edu; bakolas@austin.utexas.edu}}
}

\author{
  Vrushabh Zinage\\
  Graduate Student \\
  University of Texas at Austin \\
 Austin. Texas, USA\\
  \texttt{vrushabh.zinage@utexas.edu} \\
   \And
  Efstathios Bakolas\\
  Associate Professor \\
  University of Texas at Austin \\
 Austin. Texas, USA\\
  \texttt{bakolas@austin.utexas.edu} \\
}

\begin{document}
\maketitle

\begin{abstract}
Learning and synthesizing stabilizing controllers for unknown nonlinear control systems is a challenging problem for real-world and industrial applications. Koopman operator theory allows one to analyze nonlinear systems through the lens of linear systems and nonlinear control systems through the lens of bilinear control systems. The key idea of these methods lies in the transformation of the coordinates of the nonlinear system into the Koopman observables, which are coordinates that allow the representation of the original system (control system) as a higher dimensional linear (bilinear control) system. However, for nonlinear control systems, the bilinear control model obtained by applying Koopman operator based learning methods is not necessarily stabilizable. Simultaneous identification of stabilizable lifted bilinear control systems as well as the associated Koopman observables is still an open problem. In this paper, we propose a framework to construct these stabilizable bilinear models and identify its associated observables from data by simultaneously learning a bilinear Koopman embedding for the underlying unknown control affine nonlinear system as well as a Control Lyapunov Function (CLF) for the Koopman based bilinear model using a learner and falsifier. Our proposed approach thereby provides provable guarantees of asymptotic stability for the  Koopman based representation of the unknown control affine nonlinear control system as a bilinear system. Numerical simulations are provided to validate the efficacy of our proposed class of stabilizing feedback controllers for unknown control-affine nonlinear systems.
\end{abstract}


\section{Introduction}
%
Recently, Koopman operator techniques have proven to be powerful tools for the analysis and control of nonlinear systems whose dynamics may not be known a priori. The key idea of such methods is to associate a nonlinear system (nonlinear control system) with a linear system (bilinear control system) of higher dimension than the original system. The higher dimensional state spaces of these ``lifted'' linear or bilinear control systems are spanned by functions of states known as \textit{Koopman observables}. Unlike linearization techniques, Koopman operator methods provide higher dimensional (lifted) linear or bilinear state space models which explicitly account for nonlinearities in the original system dynamics and their validity is not limited to a small neighborhood around a reference point or trajectory as in standard linearization approaches. However, since the Koopman operator is infinite-dimensional the resulting lifted state space models will be also infinite-dimensional and consequently, the control design can become a complex, if not computationally intractable, task. In order to improve computational tractability, recent approaches in the field aim at characterizing a finite approximation of the Koopman operator via data-driven methods such as the Extended Dynamic Mode Decomposition (EDMD). EDMD mainly uses time series data to form a higher dimensional linear (bilinear) state space model that approximates the unknown nonlinear system (control-affine nonlinear control system). In addition, the connection of EDMD with Koopman operator theory has been explored in
 \cite{chen2012variants_koopman_edmd_connection_1} and extended to non-sequential time series data in \cite{tu2013dynamic_koopman_edmd_connection_2}. Further \cite{korda2018convergence} shows the convergence in the strong operator topology of the Koopman operator computed via EDMD \cite{williams2016extending_edmd} to the actual Koopman operator as the number of data points and the number of observables tend to infinity. Koopman operator theory (KOT) has been applied to robotics applications \cite{abraham2019active_robotics_1,bruder2020data_robotics_2,mamakoukas2019local_robotics_3,mamakoukas2021derivative_robotics_4,zinage2021koopman_quad}, power grid stabilization \cite{korda2018power_grid_koopman,susuki2013nonlinear_power_systems_koopman_1}, state estimation \cite{surana2017koopman_state_estimation_1,netto2018robust_koopman_state_estimation_2}, control synthesis \cite{choi2021convex_koopman_control_synthesis_1,folkestad2020data_control_synthesis_2,goswami2021bilinearization_control_synthesis_3,huang2018feedback_control_synthesis_4}, actuator and sensor placement \cite{sinha2016operator_sensor_placement}, aerospace applications \cite{zinage2021far_aero,zinage2021koopman_dual_quaternions}, analysis of climate, fluid mechanics and control of PDEs, to name but a few. Koopman operator theory postulates that a nonlinear uncontrolled system can be lifted to an equivalent (infinite-dimensional) linear system whereas a nonlinear control system to a bilinear control system. The major challenge in realizing this lifting process is that the Koopman observables are unknown and their characterization can be a complex task. \cite{lusch2018deep_koopman} presents a deep learning framework for learning the Koopman observables for uncontrolled dynamical systems. However, the learned Koopman operator is not guaranteed to be stable. \cite{fan2021learning_stable_koopman_embeddings} proposes a method to guarantee stability by learning a stable Koopman operator by utilizing a particular operator parameterization that ensures that the computed Koopman operator is Schur stable. However, the applicability of \cite{fan2021learning_stable_koopman_embeddings} is restricted to uncontrolled systems. Further,  \cite{haseli2020fast,haseli2021learning,haseli2021parallel} propose data-driven approaches for identification of Koopman invariant subspaces whose applicability is, however, limited to uncontrolled nonlinear systems.

There has been a wide interest in control design methods which are based on neural networks. Most of the proposed approaches in the relevant literature use reinforcement learning. However, there are no theoretical guarantees that the designed control system is stabilizable \footnote{A system is stabilizable if there exists a feedback controller that can render the closed-loop system asymptotically stable} which is crucial especially for safety critical applications. Towards this aim, the notion of Control Lyapunov Function (CLF) from control theory can play a vital role in characterizing the stability properties of nonlinear control systems and designing stabilizing controllers (that is, controllers that guarantee closed-loop stability). CLFs were first studied by Sontag~\cite{sontag1989universal} and Artstein \cite{artstein1983stabilization}. The existence of a CLF provides necessary and sufficient conditions for closed-loop stability of nonlinear control systems and can be viewed as a stability or safety certificate for such systems. In the control literature, there exist many approaches for the computation of Lyapunov functions for nonlinear control affine systems based on, for instance, sum-of-squares (SOS) and semidefinite programming (SDP) optimization  \cite{chesi2009guest_control_lyap_1,henrion2005positive_control_lyap_2,jarvis2003some_control_lyap_3,majumdar2017funnel_control_lyap_4}. However, these methods do not typically scale well and their applicability is limited to polynomial control-affine systems.
%
Further, \cite{ahmadi2011globally_no_polynominal_lyapfunction} showed that for a particular dynamical system, there does not exist any polynomial Lyapunov function despite the dynamical system being globally asymptotically stable.

Finding a Lyapunov function for a nonlinear system is in general a challenging task. Recently, the so-called Lyapunov neural networks methods have been proposed to learn a valid Lyapunov function that will guarantee closed-loop stability of nonlinear systems. The candidate Lyapunov functions are parameterized by means of feedforward neural networks and the Lyapunov conditions are imposed as soft constraints in the learning (optimization) problem. These methods are motivated by the fact that any continuous function can be approximated by means of a neural network with a finite number of parameters that must be learned \cite{cybenko1989approximation_neural_1,hornik1993some_neural_2}. A continuously differentiable function corresponds to a Lyapunov function for a nonlinear system if it satisfies certain conditions, which we refer to as Lyapunov conditions. One can verify whether a function learned by a neural network satisfies these conditions or not by utilizing techniques that can check certain properties of the outputs of the neural network. These verification techniques can be broadly classified into two main methods, one in which the verification is inexact and is carried out by solving a relaxed convex problem and another one in which the verification is exact and based on Mixed Integer Programming (MIP) solvers and Sequential Modulo Theories (SMT) solvers. In \cite{chang2020neural,mehrjou2020neural_lyapunov_redesign,ravanbakhsh2019learning_lyapunov_from_demonstrations,jin2020neural_control_policy} Lyapunov control methods are proposed based on Lyapunov neural networks. \cite{abate2020lyapunov_formal_synthesis} proposes formal synthesis methods for learning Lyapunov functions. However, the approaches proposed in~ \cite{chang2020neural,mehrjou2020neural_lyapunov_redesign,ravanbakhsh2019learning_lyapunov_from_demonstrations,jin2020neural_control_policy,abate2020lyapunov_formal_synthesis} assume that the nonlinear dynamics are known. \cite{dai2021lyapunov_stable_nn_control} proposes a framework for learning a Lyapunov function where the dynamics are not known. \cite{boffi2020learning_stability_from_data} provides stability certificates via a learned Lyapunov function using trajectory data only. \cite{richards2018lyapunov} proposes a framework for discrete-time polynomial (nonlinear) systems and learns a safe region of attraction (ROA) using neural networks. However, these approaches are restricted to learning linear feedback controllers or neural network based feedback controllers and do not guarantee the existence of a stabilizing feedback controller or propose a systematic method to characterize it. Further, there are no tools to analyze the stability properties of unknown {control-affine} nonlinear systems. In contrast with the aforementioned references, in this work we utilize the Koopman operator theory framework to describe, analyze and control the behavior of any known or unknown {control-affine} nonlinear system via a higher dimensional (lifted) bilinear control system.

\textbf{Contributions: } 
The main contribution of our work is four-fold. First, we propose a deep learning framework for simultaneously learning a stabilizable bilinear (lifted) state space model and the Koopman observables from data obtained from the open-loop trajectories of the latter system generated by random control inputs applied to the unknown {control-affine} nonlinear system. In our approach, closed-loop stability is guaranteed when our method can successfully learn a Control Lyapunov Function (CLF). Second, based on the learned CLF, we design a feedback controller using the celebrated universal Sontag's formula \cite{sontag1989universal} that guarantees closed-loop asymptotic stability. 
Third, our approach allows us to utilize state-of-the-art tools for verification based on SMT solvers even for the case in which the nonlinear dynamics is unknown by computing a data-driven (lifted) bilinear system (approximation of the unknown system) via KOT based methods.
Last, in contrast to recent methods
\cite{chang2020neural,mehrjou2020neural_lyapunov_redesign,ravanbakhsh2019learning_lyapunov_from_demonstrations,jin2020neural_control_policy,abate2020lyapunov_formal_synthesis,dai2021lyapunov_stable_nn_control}, which either restrict themselves to the class of linear feedback controllers or learn nonlinear feedback controllers represented by  neural networks which do not offer guarantees of closed-loop stability in general, our method ensures that the computed feedback controller is a stabilizing one which which can asymptotically steer the system to the origin. To the best knowledge of the authors, this is the first paper that proposes a method that simultaneously learns the observables together with a stabilizable Koopman Bilinear Form (KBF) which allows us to design stabilizing feedback controllers for the Koopman bilinear system. To illustrate and also validate the ability of our proposed class of stabilizing feedback controllers to steer nonlinear systems with unknown dynamics to the desired final state, we present numerical experiments for a nonlinear control system used in practical applications as well as a popular academic example.

\textbf{Organization: } The rest of the paper is organized as follows. Section \ref{sec:prelim} introduces the preliminaries followed by the problem statement. Section \ref{section:learning} discusses our approach to solve the problem. Section \ref{sec:results} discusses the results followed by some concluding remarks in Section 
\ref{sec:conclusion}.

\section{Preliminaries and Problem Statement\label{sec:prelim}}
Consider a control-affine nonlinear system given by
\begin{align}
    \dot{\boldsymbol{x}} = f(\boldsymbol{x})+g(\boldsymbol{x})\boldsymbol{u}=f(\boldsymbol{x})+\sum_{i=1}^mg_i(\boldsymbol{x}){u}_i,\quad \boldsymbol{x}(0)=\boldsymbol{x}_0
    \label{eqn:nonlinear_system}
\end{align}
where $f:\mathbb{R}^n \rightarrow \mathbb{R}^n$ is a Lipschitz continuous function, $g=[g_1,\;g_2\ldots g_m]$, where $g_i: \mathbb{R}^n \rightarrow \mathbb{R}^n$ for all $i\in \{1,\dots,m\}$, is assumed to be continuously differentiable, $\boldsymbol{x}\in\mathcal{X}\subset\mathbb{R}^n$ is the state of the system and $\boldsymbol{u}=[u_1,\;u_2,\dots,u_m]^\mathrm{T}\in\mathcal{U}\subset\mathbb{R}^m$ is the control input applied to the system where $\mathcal{X}$ and $\mathcal{U}$ are compact sets (the assumption on compactness of $\mathcal{X}$ and $\mathcal{U}$ is made for learning purposes given that it would be practically impossible to solve the learning problem over unbounded admissible input or state sets; the reader should not perceive this assumption as an indication of studying problems with state or input constraints). The function $f$ is commonly known as the drift term (or vector field) whereas $g$ is known as the actuation effect (or control vector field). We assume that the origin $\boldsymbol{x} = 0$ is the unique equilibrium of the uncontrolled system $\dot{\boldsymbol{x}}=f(\boldsymbol{x})$ in $\mathcal{X}$ (in other words, $\boldsymbol{x} = 0$ is the unique solution to the equation: $f(\boldsymbol{x})=0$ in $\mathcal{X}$).
Further, we assume that $f(\boldsymbol{x})$ and $g(\boldsymbol{x})$ are unknown in general. Next, we consider a discrete-time nonlinear control system which is obtained from the continuous-time system \eqref{eqn:nonlinear_system} after using a fourth order Runge-Kutta method:
\begin{align}
    \boldsymbol{x}_{k+1}=h(\boldsymbol{x}_k,\boldsymbol{u}_k),\quad \boldsymbol{x}_1(0)=\boldsymbol{x}_0
    \label{eqn:discretize_of_nonlinear_system}
\end{align}
where $\boldsymbol{x}_k\in\mathcal{X}$, $h:\mathcal{X}\rightarrow\mathbb{R}^n$. This discrete-time dynamical system will be used for construction of a dataset which would be used for training of the neural network (Section \ref{section:learning}).
\subsection{Koopman operator theory}
In 1931, B. O. Koopman proved the existence of an infinite dimensional linear operator that can describe the evolution of functions of states of a nonlinear system, which are known as the observables \cite{koopman1931hamiltonian}. Formally, let $\mathcal{F}$ be the space of functions spanned by the observables $\Phi:\mathbb{R}^{n}\rightarrow \mathbb{R}^N$, where $\Phi=[\phi_1,\;\phi_2,\dots,\phi_N]^\mathrm{T}$ and $N>n$, then the Koopman operator  ${\mathcal{K}}:\mathcal{F}\rightarrow\mathcal{F}$ is a linear infinite-dimensional operator that acts on functions $\Phi \in \mathcal{F}$ and is defined as follows:
\begin{align}
    {\mathcal{K}}[\Phi(\boldsymbol{x})]= \Phi\circ \mathcal{M}_t(\boldsymbol{x}),
\end{align}
where $\mathcal{M}_t$ denotes the flow map of the uncontrolled nonlinear dynamics $\dot{\boldsymbol{x}}=f(\boldsymbol{x})$ which is given by
\begin{align}
    \mathcal{M}_t\left(\boldsymbol{x}\left(t_0\right)\right)=\boldsymbol{x}\left(t_{0}\right)+\int_{t_{0}}^{t_{0}+t} {f}(\boldsymbol{x}(\tau)) \mathrm{d} \tau.
\end{align}
It can be easily verified that ${\mathcal{K}}$ is a linear operator, that is, ${\mathcal{K}}[k_1\Phi_1+k_2\Phi_2]=k_1{\mathcal{K}}[\Phi_1]+k_2{\mathcal{K}}[\Phi_2]$ for all $k_1,k_2\in\mathbb{R}$ and $\Phi_1,\;\Phi_2\in\mathcal{F}$. In practice, the Koopman operator is approximated by a finite-dimensional (truncated) operator which is subsequently used for modelling, analysis and control design. In this paper, we denote by ${\tilde{\mathcal{K}}}$ the finite-dimensional approximation (or truncation) of the Koopman operator $\mathcal{K}$ which can be represented by a matrix (we will use the same symbol for the latter operator and its matrix representation).

For control-affine nonlinear systems described by \eqref{eqn:nonlinear_system}, the time derivative of $\Phi$ along the system trajectories is given by 
\begin{align}
    \dot{\Phi}(\boldsymbol{x})&=\nabla\Phi(\boldsymbol{x})[f(\boldsymbol{x})+g(\boldsymbol{x})\boldsymbol{u}]=\nabla\Phi(\boldsymbol{x})f(\boldsymbol{x})+\nabla\Phi(\boldsymbol{x})g(\boldsymbol{x})\boldsymbol{u}\nonumber\\
    &={\tilde{\mathcal{K}}}\Phi(\boldsymbol{x}) + \nabla\Phi(\boldsymbol{x})\sum_{i=1}^mg_i(\boldsymbol{x}){u}_i.
    \label{eqn:continous_bilinear_system}
\end{align}
 We assume that $\nabla\Phi(\boldsymbol{x})g_i(\boldsymbol{x})$ belongs to the span of $\Phi(\boldsymbol{x})$. In other words, there exists a constant matrix $Q_i$ such that
 \begin{align}
    \nabla\Phi(\boldsymbol{x})g_i(\boldsymbol{x})=Q_i\Phi(\boldsymbol{x}). \nonumber
 \end{align}
 This is a reasonable assumption to make given that, as is shown in~\cite{bruder2021advantages_bilinear}, for sufficiently large number of Koopman observables, the system governed by \eqref{eqn:nonlinear_system} can be equivalently modelled as a Koopman Bilinear Form (KBF) as follows \cite{butcher1987numerical}:
 \begin{align}
     \dot{\boldsymbol{z}}={\tilde{\mathcal{K}}}\boldsymbol{z}+\sum_{i=1}^mQ_i\boldsymbol{z}u_i,
     \label{eqn:continous_bilinear_system}
 \end{align}
where $\boldsymbol{z}:=\Phi(\boldsymbol{x})$ and $\boldsymbol{z}\in\mathcal{Z}$ where $\mathcal{Z} =\{\boldsymbol{z}=\Phi(\boldsymbol{x}):\boldsymbol{x}\in\mathcal{X}\}$. Note that $\boldsymbol{z}=0$ is a equilibrium point for the bilinear system \eqref{eqn:continous_bilinear_system}. After applying Euler discretization to the KBF and assuming zero order hold, we get 
 \begin{align}
     \boldsymbol{z}_{k+1}=\boldsymbol{z}_k + T\tilde{\mathcal{K}}\boldsymbol{z}_k+T\sum_{i=1}^mQ_i\boldsymbol{z}_ku_i=\tilde{\mathcal{K}}_d\boldsymbol{z}_k+\sum_{i=1}^mB_i\boldsymbol{z}_ku_i,
 \end{align}
where $T$ is the sampling period, $\tilde{\mathcal{K}}_d := I+T\tilde{\mathcal{K}}$, $I$ is the identity matrix and $B_i=T Q_i$. Note that the local truncation error, $ \| \boldsymbol{z}(t_k) - \boldsymbol{z}_k\|$, satisfies the following upper bound:
\begin{align}
    \| \boldsymbol{z}(t_k) - \boldsymbol{z}_k\|\leq pLT^2,
\end{align}
where $L$ and $p$ are given by
\begin{align}
    L:=\Big\|\tilde{\mathcal{K}}_d+\sum_{i=1}^m B_i\boldsymbol{z}_k {u}_{i}\Big\|,\quad p:=L \max _{k T \leq t<k T+T}\|\boldsymbol{z}(t)\|.
\end{align}
Note that the truncation error tends to zero as $T$ tends to zero. Hence, even if for a given $T$, the truncation error between the states of the discrete-time system and the continuous-time system is not sufficiently small, $T$ can be reduced accordingly. Note that in contrast to discretization approaches like Runge-Kutta, the Euler discretization preserves the bilinear form in the discrete-time state space model. This was the main motivation as to why Euler discretization was chosen over the Runge-Kutta methods.

\begin{remark}
\normalfont Of particular interest is the case in which one can find a set of Koopman observables such that the unknown nonlinear system can be represented by a (continuous-time) linear time invariant (LTI) system as follows:
\begin{align}\label{eq:LTI}
    \dot{\boldsymbol{z}}=A\boldsymbol{z}+B\boldsymbol{u}.
\end{align}
The continuous-time LTI in \eqref{eq:LTI} can be associated with the following discrete-time LTI system:
\begin{align}
        {\boldsymbol{z}_{k+1}}=A_d\boldsymbol{z}_k+B_d\boldsymbol{u}_k,
        \label{eqn:linear_koopman}
\end{align}
where $A_d=e^{AT}$, $B_d={\int_0^Te^{At}\mathrm{d}t}B$ and $T$ is the sampling period. Recently, the approach of lifting a nonlinear control system to a linear control system has gained a lot of attention \cite{abraham2019active_robotics_1} due to the availability of rich libraries of tools to analyze and design controllers for linear systems. However, lifting a nonlinear system into a higher dimensional linear system can be quite restrictive in practice because it may be hard to find a linear system that accurately describe the behavior of the original nonlinear system over a large portion of the state space, as pointed out in \cite{brunton2016koopman_linear_limitation}. Further, the realization of the lifting process based on Koopman operator methods applied to a control-affine nonlinear system yields a bilinear control system rather than a linear system. This (lifted) bilinear representation of the control-affine nonlinear system has several advantages over the counterpart linear system representation as pointed out in \cite{bruder2021advantages_bilinear}.
\end{remark}
\smartdiagramset{ text width=4.5cm, font=\fontsize{10pt}{9pt}\selectfont, uniform color list=white!30 for 3 items,
uniform arrow color=true, arrow color=gray!30!black, arrow line width=2pt, module x sep=6.4cm, module minimum height=2.7cm}
\begin{figure*}
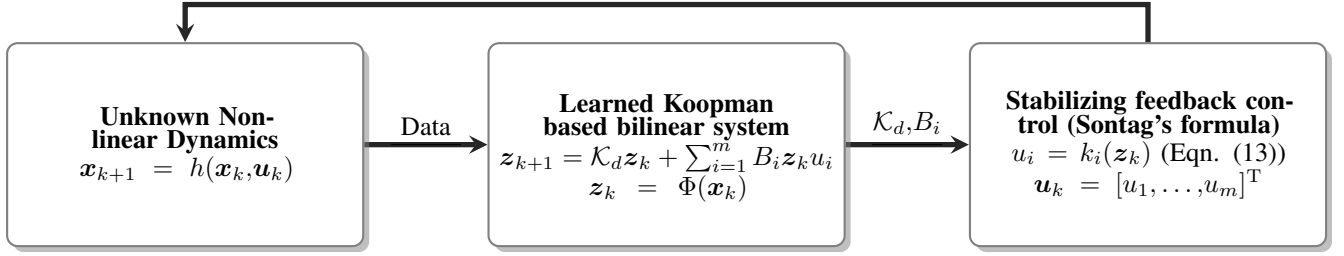

\centering
    \smartdiagramadd[flow diagram:horizontal]{\textbf{Unknown Nonlinear Dynamics}\\ $\boldsymbol{x}_{k+1}=h(\boldsymbol{x}_{k}{,}\boldsymbol{u}_{k})$,\\
  \textbf{Learned Koopman based bilinear system}\\
  $\boldsymbol{z}_{k+1}=\mathcal{K}_d\boldsymbol{z}_k+\sum_{i=1}^mB_i\boldsymbol{z}_ku_i$\\
  $\boldsymbol{z}_k=\Phi(\boldsymbol{x}_k)$, \textbf{Stabilizing feedback control (Sontag's formula)}\\
  ${{u}}_i=k_i(\boldsymbol{z}_k)$ (Eqn. \eqref{eqn:sontag_formula})\\
  $\boldsymbol{u}_k=[u_1{,}\dots{,}u_m]^\mathrm{T}$
}{}
 \smartdiagramconnect{shorten >=10pt,shorten <=10pt,"Data"
{text=black}}{module1/module2}
 \smartdiagramconnect{shorten >=10pt,shorten <=10pt,"$\mathcal{K}_d{,} B_i$"
{text=black}}{module2/module3}
 \caption{ Learning and control framework for unknown nonlinear dynamics using Koopman operator theory}
 \label{fig:feedback_diagram}
\end{figure*}
\subsection{Existence of stabilizing feedback controllers}
Given the bilinear system  \eqref{eqn:continous_bilinear_system}, let us define the CLF as follows:
\begin{definition}
\normalfont A Control Lyapunov Function is a continuously differentiable positive definite function $V:\mathcal{D}\rightarrow \mathbb{R}_{+}$, where $0 \in \mathcal{D}$ (that is, $V$ is positive everywhere in $\mathcal{D}$ expect at $\boldsymbol{z}=0$ where it is zero) such that the infimum of the Lie derivative of $V$ over all inputs is negative. More precisely,
\begin{subequations}
\begin{align}
    \underset{\boldsymbol{u}\in\mathcal{U}}{\text{inf}}\;\;&\dot{V}(\boldsymbol{z})<0\quad\\
    \text{where}\quad& \dot{V}(\boldsymbol{z}) :=\frac{\partial V}{\partial \boldsymbol{z}}\dot{\boldsymbol{z}}=\frac{\partial V}{\partial \boldsymbol{z}}\tilde{\mathcal{K}}\boldsymbol{z}+\frac{\partial V}{\partial \boldsymbol{z}}\sum_{i=1}^mQ_i\boldsymbol{z}u_i.
\end{align}
\label{eqn:clf}
\end{subequations}
\end{definition}
If the infimum of the Lie derivative of the CLF over all $\boldsymbol{u}\in\mathcal{U}$ is negative, then there exists a control input for which $\dot{V}(\boldsymbol{z})$ is negative along the trajectories of the closed-loop system. In particular, it can be shown that in the latter case there exists a feedback controller $\boldsymbol{u} :=  k(\boldsymbol{z}) = [k_1(\boldsymbol{z}),\dots,k_m(\boldsymbol{z})]^\mathrm{T}$ that renders the closed-loop system asymptotically stable (in other words, $k(\boldsymbol{z})$ is a stabilizing feedback controller). This is stated formally as follows:

\begin{theorem}
\normalfont \cite{sontag1983lyapunov} For the system given by \eqref{eqn:continous_bilinear_system}, there exists a continuously differentiable function $k(\boldsymbol{z})$ for all $\boldsymbol{z}\in\mathbb{R}^N\setminus\{0\}$ and continuous at $\boldsymbol{z}=0$ such that the controller $\boldsymbol{u}=k(\boldsymbol{z})$ renders the zero solution $\boldsymbol{z}=0$ of the closed-loop system asymptotically stable if and only if 
there exists a Control Lyapunov Function (CLF) $V(\boldsymbol{z})$ such that

    (C1) For all $\boldsymbol{z}\neq 0$, $\sum_{i=1}^m\frac{\partial V}{\partial \boldsymbol{z}}Q_i\boldsymbol{z}u_i=0$ implies $\frac{\partial V}{\partial \boldsymbol{z}}\tilde{\mathcal{K}}\boldsymbol{z}<0$
    
    (C2) For all $\epsilon>0$, there exists $\delta>0$ such that $\|\boldsymbol{z}\|<\delta$ implies the existence of $\|\boldsymbol{u}\|<\epsilon$ satisfying \eqref{eqn:clf}

The condition (C2) is also known as the small control property.
If both conditions (C1) and (C2) hold true, then the feedback controller $\boldsymbol{u} := k(\boldsymbol{z})=[k_1(\boldsymbol{z}), k_2(\boldsymbol{z}), \dots, k_m(\boldsymbol{z})]^\mathrm{T}$, where the $i$-th component $k_i(\boldsymbol{z})$ of the feedback controller $k(\boldsymbol{z})$ satisfies the universal Sontag's formula \cite{sontag1989universal}:
\begin{align}
        k_i(\boldsymbol{z})=\begin{cases}-\frac{\boldsymbol{c}_i(\boldsymbol{z})[a(\boldsymbol{z})+\sqrt{a^2(\boldsymbol{z})+\sigma^2(\boldsymbol{z})}]}{\sigma(\boldsymbol{z})},\qquad\text{if}~ \sigma(\boldsymbol{z})\neq 0\\
    0,\qquad\qquad\qquad\qquad \qquad \qquad \text{otherwise}
    \end{cases}
    \label{eqn:sontag_formula}
\end{align}
where $a(\boldsymbol{z})=\frac{\partial V}{\partial \boldsymbol{z}}\tilde{\mathcal{K}}\boldsymbol{z}$, $\sigma(\boldsymbol{z})=\sum_{i=1}^m(\frac{\partial V}{\partial \boldsymbol{z}}Q_i\boldsymbol{z}{u}_i)^2$, and  $\boldsymbol{c}_i(\boldsymbol{z})=\frac{\partial V}{\partial \boldsymbol{z}}Q_i\boldsymbol{z}$, will be  a stabilizing controller (in other words, the controller $k(\boldsymbol{z})$ will render the closed-loop system asymptotically stable)
\label{theorem:sontags}
\end{theorem}
\begin{remark}
\normalfont In contrast to recent methods which compute linear feedback controllers \cite{chang2020neural,jin2020neural_control_policy} or  nonlinear feedback controllers represented by neural networks \cite{dai2021lyapunov_stable_nn_control} without offering any guarantees for closed-loop stability in general, our approach guarantees the existence of a stabilizing controller that will asymptotically steer the state of the Koopman based bilinear system \eqref{eqn:nonlinear_system} to the origin.
\end{remark}

\subsection{Problem statement}
Next, we provide the precise formulation of the problem addressed in this paper:
\begin{problem}
\normalfont Given the data snapshots $X$ from the unknown nonlinear control system \eqref{eqn:nonlinear_system}, compute the Koopman observables $\boldsymbol{z}=\Phi(\boldsymbol{x})$ and the matrices $[\tilde{\mathcal{K}}_d,\; B_1,\; \dots,\; B_{m}]$ governing the Koopman based bilinear system \eqref{eqn:continous_bilinear_system}. Further, design a  feedback controller that renders the zero solution of the Koopman based bilinear system \eqref{eqn:continous_bilinear_system}
\end{problem}
In the following sections, we explain in detail how our learning framework simultaneously learns the lifted bilinear system and a valid CLF, thereby consequently allowing us to design stabilizing feedback controllers for the lifted bilinear system.

\section{Learning a stabilizable bilinear control system using Koopman operator theory\label{section:learning}}

\begin{figure}[]
 \captionsetup[subfigure]{justification=centering}
 \centering
 \begin{subfigure}{0.5\textwidth}
{\includegraphics[width=7.6cm]{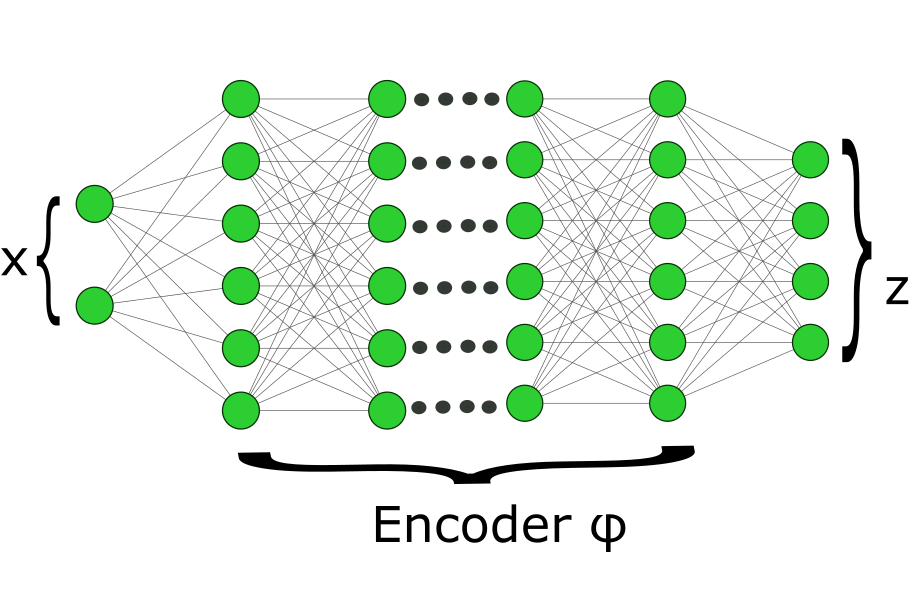}}
\caption{An encoder $\Phi(\boldsymbol{x})$ which lifts the state $\boldsymbol{x}$ to a higher dimensional state space $\boldsymbol{z}$}
\label{fig:encoder}
 \end{subfigure}
 \begin{subfigure}{0.36\textwidth}
 \includegraphics[width=7cm]{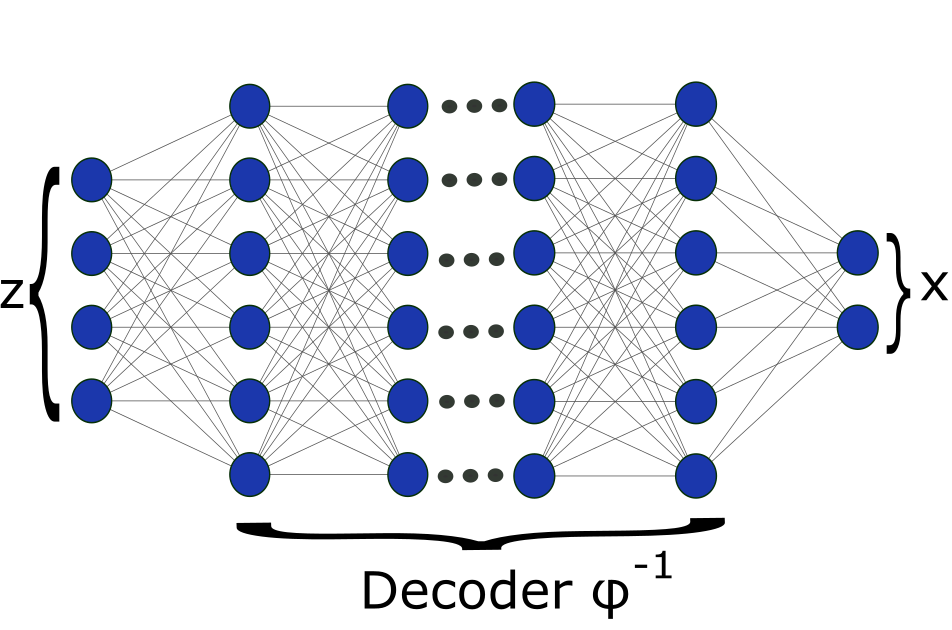}
 \caption{A decoder $\Phi^{-1}(\boldsymbol{z})$ which brings back the lifted state $\boldsymbol{z}$ to the original state $\boldsymbol{x}$}
\label{fig:decoder}
 \end{subfigure}
 \begin{subfigure}{0.52\textwidth}
 \includegraphics[width=8.3cm]{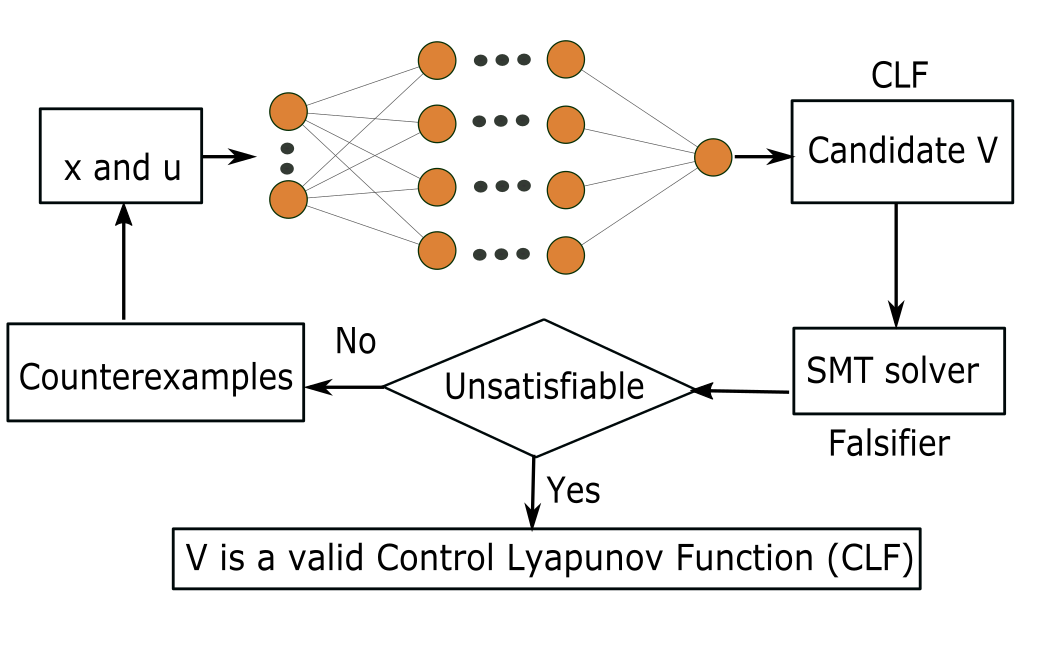}
\caption{Training a Lyapunov neural network to learn a Control Lyapunov Function (CLF) that ensures the existence of a stabilizing controller for the bilinear system (Theorem \ref{theorem:sontags})}
\label{fig:lyapunov}
\end{subfigure}
 \caption{Deep learning framework to learn a stabilizable bilinear control system and a Control Lyapunov Function (CLF)}
\label{fig:learning_framework}
\end{figure}

In this section, we present a learning approach to simultaneously learn the Koopman observables and a valid CLF for the learned bilinear system. Let the state snapshots $\{\boldsymbol{x}_k\}_{k=1}^{N_d}$ and the corresponding control inputs $\{\boldsymbol{u}_k\}_{k=0}^{N_d-1}$ be obtained when the control input $\boldsymbol{u}_k$ is applied to the discrete-time system \eqref{eqn:discretize_of_nonlinear_system}  \eqref{eqn:discretize_of_nonlinear_system} to transfer the state from $\boldsymbol{x}_k$ to $\boldsymbol{x}_{k+1}$ for all $k\in\{1,2\dots,N_d-1\}$ and let $N_d$ denote the total number of data snapshots. Consider an encoder $\boldsymbol{z}=\Phi(\boldsymbol{x}; \theta):\mathbb{R}^n\rightarrow\mathbb{R}^N$ which maps the state $\boldsymbol{x}\in\mathbb{R}^n$ to a higher dimensional lifted state $\boldsymbol{z}\in\mathbb{R}^N$ where $N>n$ and $\theta$ denotes the vector of parameters of the neural network. Similarly, let $\boldsymbol{x}=\Phi^{-1}(\boldsymbol{z};\theta):\mathbb{R}^N\rightarrow\mathbb{R}^n$ denote the decoder which maps the lifted state back to the original state $\boldsymbol{x}$ as shown in Fig. \ref{fig:learning_framework}. For notational simplicity, we represent $\Phi^{-1}(\boldsymbol{z};\theta)$ by $\Phi^{-1}(\boldsymbol{z})$. We construct a CLF $V(\boldsymbol{z};\theta)$, to be the output of a feedforward neural network. The main motivation for using feedforward neural networks for representing CLF is that they are expressive in the sense that any continuous function can be represented by means of a feedforward network with a finite number of parameters. For notational simplicity, we denote $V(\boldsymbol{z};\theta)$ by $V_\theta(\boldsymbol{z})$. Since the CLF has to be continuously differentiable, a smooth $\texttt{tanh}$ activation function is used. One can also use the smooth approximations of the $\texttt{ReLU}$ activation function. The objective is to simultaneously learn a valid CLF, Koopman observables $\Phi$, the discrete-time Koopman operator $\tilde{\mathcal{K}}_d$ and the matrices $B_i$ for all $i\in\{1,2,\dots m\}$ that appear in the governing equations of the bilinear system by minimizing the following loss function $\mathcal{L}$ given by

\begin{align}
    \mathcal{L}=\alpha_1\mathcal{L}_{\text{recons}}+\alpha_2\mathcal{L}_{\text{dyn}}+\alpha_3\mathcal{L}_{\text{lyap}}+\alpha_4\mathcal{L}_\text{ROA}
    \label{eqn:total_loss}
\end{align}
where $\alpha_1$, $\alpha_2$, $\alpha_3$ and $\alpha_4$ are positive hyperparameters. Next, we describe the individual losses $\mathcal{L}_{\text{recons}}$, $\mathcal{L}_{\text{dyn}}$, $\mathcal{L}_{\text{phy}}$, $\mathcal{L}_{\text{lyap}}$, $\mathcal{L}_\text{ROA}$ whose weighted sum constitute the overall loss function $\mathcal{L}$.

\noindent $\mathrm{i)}~$\textbf{Reconstruction loss $\mathcal{L}_{\text{recons}}$}: The reconstruction loss ensures that the encoder is able to lift the state $\boldsymbol{x}$ and the decoder is able to project back the lifted state $\boldsymbol{z}$ to $\boldsymbol{x}$. The expression for $\mathcal{L}_{\text{recons}}$ is given by
\begin{align}
    \mathcal{L}_{\text{recons}}=\frac{1}{N_d}\sum_{k=1}^{N_d}\|\boldsymbol{x}_k-\Phi^{-1}(\Phi(\boldsymbol{x}_k))\|^2_2\nonumber
\end{align}

\noindent $\mathrm{ii)}$~\textbf{Bilinear control system loss $\mathcal{L}_{\text{dyn}}$:} The dynamical system loss $\mathcal{L}_{\text{dyn}}$ (also known as the bilinear control system loss in this case) represents the extent to which the observables obey the governing Koopman based bilinear system and is given by the following expression
$$
\mathcal{L}_{\text{dyn}}=\frac{1}{N_d-1} \sum_{i=1}^{N_d-1}\Big\|{\Phi}\left(\boldsymbol{x}_{i+1}\right)-\tilde{\mathcal{K}}_d {\Phi}\left(\boldsymbol{x}_{i}\right)-\sum_{j=1}^{{m}}[\boldsymbol{u}_{i}]_j B_j {\Phi}\left(\boldsymbol{x}_{i}\right)\Big\|^2_2
$$
During every epoch, the matrices $\tilde{\mathcal{K}}_d, B_1, \dots, B_{m}$ are updated as follows:
\begin{align}
\left[\begin{array}{lll}
\tilde{\mathcal{K}}_d & B_1 \dots & B_{m}
\end{array}\right]=\beta_{N_d} \Psi_{N_d}^{T}\left(\Psi_{N_d} \Psi_{N_d}^{T}\right)^{-1}
\label{eqn:update_matrices}
\end{align}
where the matrices $\Psi_{N_d}$ and $\beta_{N_d}$ are given by

\begin{align}
&\Psi_{N_d}=\left[\begin{array}{lll}
\left(\begin{array}{c}
1 \\
\boldsymbol{u}_{1}
\end{array}\right) \otimes\Phi(\boldsymbol{x}_{1}), & \cdots &, \left(\begin{array}{c}
1 \\
\boldsymbol{u}_{N_d-1}
\end{array}\right) \otimes\Phi(\boldsymbol{x}_{N_d-1})
\end{array}\right],\nonumber\\
&\beta_{N_d}=[\Phi(\boldsymbol{x}_{2}),\;\dots,\Phi(\boldsymbol{x}_{N_d})]
\end{align}
where $\otimes$ denotes the Kronecker product. Note that for given $\Phi$ and $\Phi^{-1}$, the matrices $\tilde{\mathcal{K}}, B_1,\cdots B_{m}$ updated as in \eqref{eqn:update_matrices} minimize $\mathcal{L}_{\text{dyn}}$ if $\Psi_{N_d}$ is a full row-rank matrix. This approach is known as the Extended Dynamic Mode Decomposition (EDMD) \cite{tu2013dynamic_koopman_edmd_connection_2}. Note that $\Psi_{N_d}$ can be made full row-rank if we increase the number of data-snapshots. Further, these matrices are updated optimally during every epoch of training the neural network. However, the major challenge is that the right Koopman observables $\Phi$ are unknown at the beginning. Hence, the loss initially is not zero and can actually be significantly large.


\noindent $\mathrm{iii)}$~\textbf{Control Lyapunov risk $\mathcal{L}_{\text{lyap}}$:} 
The control design based on CLF involves minimizing the minimax cost which is represented as follows \cite{chang2020neural}:
\begin{align}
\min_{\theta, u\in\mathcal{U}} \max _{\boldsymbol{z} \in \mathcal{Z}}\left(\max \left(0,-V_{\theta}(\boldsymbol{z})\right)+\max \left(0, \nabla V_{\theta}(\boldsymbol{z})\dot{\boldsymbol{z}}\right)+V_{\theta}^{2}(0)\right)
\label{eqn:lyapunov_analytical}
\end{align}
where $\nabla V_{\theta}(\boldsymbol{z})$ denotes the gradient of $V_{\theta}(\boldsymbol{z})$ with respect of $\boldsymbol{z}$. The time derivative of $V_{\theta}(\boldsymbol{z})$ along the system's trajectories is defined as follows:
\begin{align}
  \dot{V}_{\theta}(\boldsymbol{z}):=\nabla V_{\theta}(\boldsymbol{z})\dot{\boldsymbol{z}}=\frac{\partial V}{\partial \boldsymbol{z}}\tilde{\mathcal{K}}\boldsymbol{z}+\frac{\partial V}{\partial \boldsymbol{z}}\sum_{i=1}^m{Q_i\boldsymbol{z}u_i}=\frac{\partial V}{\partial \boldsymbol{z}}\frac{(\tilde{\mathcal{K}}_d-I)\boldsymbol{z}}{T}+\frac{\partial V}{\partial \boldsymbol{z}}\sum_{i=1}^m\frac{B_i\boldsymbol{z}u_i}{T}\nonumber
\end{align}
The first term in \eqref{eqn:lyapunov_analytical} ensures that the CLF is positive definite, the second term ensures that the Lie derivative of CLF is negative and the last terms ensure that the value of CLF at the origin is zero. The control Lyapunov loss function $\mathcal{L}_{\text{lyap}}$ measures the degree of violation of the Lyapunov conditions mentioned in \eqref{eqn:clf}.
{Let the value of $\min_{\theta, u\in\mathcal{U}} \max _{\boldsymbol{z} \in \mathcal{Z}}\left(\max \left(0,-V_{\theta}(\boldsymbol{z})\right)+\max \left(0, \nabla V_{\theta}(\boldsymbol{z})\dot{\boldsymbol{z}}\right)+V_{\theta}^{2}(0)\right)$ be $G(\boldsymbol{z}^\star,\boldsymbol{u}^\star)$. If $V_{\theta}(\boldsymbol{z})$ is a valid CLF, then $G(\boldsymbol{z}^\star,\boldsymbol{u}^\star)=G(\boldsymbol{z}_{1})=\dots =G(\boldsymbol{z}_{N_d})=0$. Since a valid CLF is not known during the training process and given the set $\mathcal{Z}=\{\boldsymbol{z}_1,\;\boldsymbol{z}_2,\dots,\boldsymbol{z}_{N_d}\}$, the conditional probability $\mathbb{P}(G(\boldsymbol{z}^\star,\boldsymbol{u}^\star)=G(\boldsymbol{z}_1,\boldsymbol{u}_1)|\mathcal{Z})=\mathbb{P}(G(\boldsymbol{z}^\star,\boldsymbol{u}^\star)=G(\boldsymbol{z}_2,\boldsymbol{u}_2)|\mathcal{Z})= \dots =\mathbb{P}(G(\boldsymbol{z}^\star,\boldsymbol{u}^\star)=G(\boldsymbol{z}_{N_d},\boldsymbol{u}_{N_d})|\mathcal{Z})$. Therefore, the optimal unbiased Monte Carlo estimate \cite{chang2020neural} of the control Lyapunov risk is given by the sample mean as follows: }
\begin{align}
    \mathcal{L}_{\text{lyap}}=\frac{1}{N_d}\sum_{i=1}^{N_d}\left(\max \left(0,-V_{\theta}(\boldsymbol{z}_i)\right)+\max \left(0, \nabla V_{\theta}(\boldsymbol{z}_i)\dot{\boldsymbol{z}}\right)+V_{\theta}^{2}(0)\right)
    \label{eqn:lyapunov_sample_mean}
\end{align}
The hard constraint $V_\theta(0)=0$ is satisfied by setting the biases of the neural network $V_\theta(\boldsymbol{z})$ to be zero. The violation of the Lyapunov conditions leads to failure in designing control inputs as these conditions need to be guaranteed over all states in $\mathcal{D}$. To avoid this issue, a first-order logic (also known as the falsification constraint) is incorporated which generates a counter-example that would not satisfy the Lyapunov conditions \eqref{eqn:clf}. In other words, the first-order logic $\mathcal{F}_{\varepsilon}(\boldsymbol{z})$ can be represented as 
\begin{align}
    \mathcal{F}_{\varepsilon}(\boldsymbol{z})=\Big(\sum_{i=1}^{N} \boldsymbol{z}_{i}^{2}= \sum_{i=1}^{N} \Phi(\boldsymbol{x})^{2}_i\geq \varepsilon>0\Big) \wedge\Big(V_\theta(\boldsymbol{z}) \leq 0 \vee \nabla V_\theta(\boldsymbol{z})\dot{\boldsymbol{z}} \geq 0\Big)
    \label{eqn:falsification_constraint}
\end{align}
\normalfont To avoid numerical sensitivity issues, the value of $\varepsilon$ is orders of magnitude smaller than the dimension of $\mathcal{Z}$. Further, $\varepsilon$ is chosen such that $\sum_{i=1}^{n} \boldsymbol{x}_{i}^{2}\geq \varepsilon$ would imply that $\sum_{i=1}^{N} \boldsymbol{z}_{i}^{2}\geq \varepsilon$. Therefore, $\varepsilon$ is chosen such that
\begin{align}
    \varepsilon \ll \min\left\{1,\;\|\mathcal{Z}\|,\;\min\left\{\sum_{i=1}^{n} \boldsymbol{x}_{i}^{2},\sum_{i=1}^{N} \boldsymbol{z}_{i}^{2}\right\}\right\}.
\end{align}
We use a Satisfiability Modulo Theories (SMT) algorithm (SMT algorithms are used to determine whether a mathematical formula is satisfiable or not) for generating counterexamples which satisfy the falsification constraint \eqref{eqn:falsification_constraint}. The problem of generating examples that satisfy the nonlinear constraints is highly non-convex and NP hard. However, recent progress among the class of SMT algorithms has shown to be effective in solving problems with such nonlinear constraints.
The neural network is trained until the SMT solver is not able to generate a counterexample satisfying the falsification constraint. Once a counterexample is generated, the training data are updated accordingly to further train the neural network. Note that a SMT solver never fails to generate counterexamples which satisfy the falsification constraint \eqref{eqn:falsification_constraint} if there are any. This is rigorously proved for SMT solvers such as dReal in \cite{gao2013dreal}.

If the dimension of the nonlinear system is large, learning a valid CLF using $\mathcal{L}_{\text{dyn}}$ and $\mathcal{F}_{\varepsilon}$ might be computational expensive for a SMT solver. To improve computational tractability, we simplify the computation of $\mathcal{L}_{\text{dyn}}$ and $\mathcal{F}_{\varepsilon}$ by considering the set of candidate CLF's as follows.
Consider a set of candidate Control Lyapunov Functions $V(\boldsymbol{z};\theta)$ as follows:
\begin{align}
    V(\boldsymbol{z};\theta)=\boldsymbol{z}^{\mathrm{T}}(\gamma I+W_N(\boldsymbol{z})^{\mathrm{T}} W_N(\boldsymbol{z}))\boldsymbol{z},
\end{align}
where $W_N(\boldsymbol{z})$ is a $n_w\times N$ matrix that corresponds to the output of a feedforward neural network, $\gamma>0$ and $n_w$ is the number of hyper-parameters. Clearly, $V(\boldsymbol{z}; \theta)$ is positive definite given that the matrix $I + W_N(\boldsymbol{z})^{\mathrm{T}} W_N(\boldsymbol{z})$ is positive definite as the sum of a positive definite matrix and a positive semi-definite matrix. Therefore, $\mathcal{L}_{\text{lyap}}$ and $\mathcal{F}_{\varepsilon}(\boldsymbol{z})$ can be written as follows:
\begin{align}
     &\mathcal{L}_{\text{lyap}}=\frac{1}{N_d}\sum_{i=1}^{N_d}\max \left(0, \nabla V_{\theta}(\boldsymbol{z}_i)\dot{\boldsymbol{z}}_i\right),\nonumber\\
    &\mathcal{F}_{\varepsilon}(\boldsymbol{z})=\left(\sum_{i=1}^{N} \boldsymbol{z}_{i}^{2}= \sum_{i=1}^{N} \Phi(\boldsymbol{x})_{i}^{2}\geq \varepsilon\right) \wedge  (\nabla V_\theta(\boldsymbol{z})\dot{\boldsymbol{z}} \geq 0)\nonumber
\end{align}
\begin{remark}
\normalfont A stabilizable Koopman based linear model can also be learned by suitably modifying $\mathcal{L}_{\text{dyn}}$ and $\mathcal{L}_{\text{lyap}}$ as
\begin{align}
  \mathcal{L}_{\text{dyn}}=\frac{1}{N_d-1}\sum_{i=1}^{N_d-1}\left\|{\Phi}\left(\boldsymbol{x}_{i+1}\right)-\tilde{\mathcal{K}}_d {\Phi}\left(\boldsymbol{x}_{i}\right)-B_d\boldsymbol{u}_i \right\|_{2}^{2}\nonumber  
\end{align}
where the matrices $A_d$ and $B_d$ (Eqn. \eqref{eqn:linear_koopman}) are updated during each epoch as follows:
\begin{align}
    [\tilde{\mathcal{K}}_d, B_d]\triangleq[A_d, B_d]=\beta_{N_d}\left[\Psi_{N_d}, \mathbf{U}\right]^{\dagger}
\end{align}
where $\dagger$ denotes the Moore-Pseudo inverse and $\beta_{N_d}$, $\Psi_{N_d}$ and $\mathbf{U}$ are given by
\begin{subequations}
\begin{align}
    \Psi_{N_d} & =[\Phi(\boldsymbol{x}_1),\dots,\Phi(\boldsymbol{x}_{N_d-1})],\quad\\
    \beta_{N_d} &=[\Phi(\boldsymbol{x}_2),\dots,\Phi(\boldsymbol{x}_{N_d})],\quad\\
    \mathbf{U}&=[\boldsymbol{u}_1,\dots,\boldsymbol{u}_{N_d}].
\end{align}
\end{subequations}
Further the control Lyapunov risk and the falsification constraint remain the same except that $\nabla V_\theta(\boldsymbol{z})$ now becomes:
\begin{align}
  \nabla V_{\theta}(\boldsymbol{z})=\frac{\partial V}{\partial \boldsymbol{z}}\frac{(\tilde{\mathcal{K}}_d-I)\boldsymbol{z}}{T}+\frac{\partial V}{\partial \boldsymbol{z}}B
\end{align}
\end{remark}

\noindent $\mathrm{iv)}$~\textbf{Region of attraction (RoA) loss function $\mathcal{L}_{\text{ROA}}$:} Let $\phi(t,\boldsymbol{z})$ be the solution to the system of ordinary differential equations which describe the dynamics of a nonlinear system with initial condition $\boldsymbol{z}$ at time $t=0$. Then, the Region of Attraction (RoA) is defined as the set of all points such that $\underset{t\rightarrow \infty}{\text{lim}}\phi(t,\boldsymbol{z})=0$ \cite{hasan_khalil}.
 
Finding the exact region of attraction either analytically or numerically is not possible for many practical cases. However, one can use Lyapunov based methods to estimate the RoA of nonlinear systems \cite{pylorof2018stabilization_pylorof_1,pylorof2016analysis_pylorof_2}. If there exists a Lyapunov function that satisfies the conditions of asymptotic
stability over a domain $\mathcal{Z}$, then the simplest estimate of RoA is given by the set $\Omega_c=\{\boldsymbol{z}\in\mathbb{R}^N:\;V(\boldsymbol{z})\leq c\}\subset \mathcal{Z}$. The RoA loss $\mathcal{L}_\text{ROA}$ is defined as follows:
 \begin{align}
     \mathcal{L}_\text{ROA}=\frac{1}{N_d}\sum_{i=1}^{N_d}\|\boldsymbol{z}_i\|_2-\gamma_4 V_\theta(\boldsymbol{z}_i)
 \end{align}
 where $\gamma_4>0$ is a tunable parameter. The loss function $\mathcal{L}_\text{ROA}$ characterized how fast the CLF increases compared to the radius of the level sets. The region of attraction is also called as the region of asymptotic stability or domain / basin of attraction~\cite{hasan_khalil}.

\begin{remark}
\normalfont Although, the optimization problem (i.e., minimizing the loss given in \eqref{eqn:total_loss}) is highly non-convex, recent results in deep learning have been successful in finding global minima for these non-convex problems. 
\end{remark}

The overall algorithm used to learn and control an unknown {control-affine} nonlinear system is described in Algorithm \ref{alg:learn_bilinear}. Its main steps can be summarized as follows. The function $\texttt{LEARNING}$ takes the data snapshots and returns the learned matrices governing the higher dimensional bilinear system, the Koopman observables (encoder), the decoder and a valid CLF. The learned CLF and the matrices governing the bilinear systems are then used to design a stabilizing feedback controller based on Sontag's formula \eqref{eqn:sontag_formula}. The function $\texttt{CONTROL}$ takes the initial state $\boldsymbol{x}_0$ and computes the control sequence $U$ (via the learned Koopman based bilinear model and the Control Lyapunov function) which steers the unknown nonlinear control system to the origin.
\begin{algorithm}[H]
 \caption{Learning Koopman operator based stabilizable bilinear model for control}
 \small
\hspace*{\algorithmicindent} \textbf{Input:} $N_d$, $n$, $m$, $T$, $X=\{\boldsymbol{x}_k,\;\boldsymbol{u}_k\}_{k=1}^{N_d}$ \\
 \hspace*{\algorithmicindent} \textbf{Output:} $\Phi$, $\Phi^{-1}$, $\mathcal{K}_d$, $B_i\;\forall\;i\in\{1,\dots,m \}$ , $\boldsymbol{u}=k(\boldsymbol{z})$
\begin{algorithmic}[1]
\State $\textbf{function}\;\;\texttt{LEARNING}(\{\boldsymbol{x}_k\}_{k=1}^{N_d},\;\{\boldsymbol{u}_k\}_{k=1}^{N_d})$
\State $\quad\textbf{Repeat:}$
\State $\quad\Phi,\;\Phi^{-1}\gets\text{NN}_{\theta}(\boldsymbol{x},\boldsymbol{z})$\Comment{Encoder and decoder}
\State $\quad V_{\theta}(\boldsymbol{z}) \leftarrow \mathrm{NN}_{\theta}(\boldsymbol{z})$\Comment{Candidate Control Lyapunov Function (CLF)}
\State $\quad \;\left[\mathcal{K}_d, \dots , \mathrm{B}_{m}
\right]\gets\beta_{N_d} \Psi_{N_d}^{T}\left(\Psi_{N_d} \Psi_{N_d}^{T}\right)^{-1}$\Comment{Update bilinear system matrices}
\State $\quad   \nabla V_{\theta}(\boldsymbol{z})\gets\frac{\partial V}{\partial \boldsymbol{z}}\frac{(\mathcal{K}_d-I)\boldsymbol{z}}{T}+\frac{\partial V}{\partial \boldsymbol{z}}\sum_{i=1}^m\frac{B_i\boldsymbol{z}u_i}{T} $
\State $\quad \text { Compute total loss } \mathcal{L}(\theta)\;\;\text{from Eqn. \eqref{eqn:total_loss}}$
\State $\quad \theta \leftarrow \theta+\alpha \nabla_{\theta} \mathcal{L}(\theta)$\Comment{Update weights}
\State $\quad\textbf{until}\; \text{convergence}$
\State $\quad\textbf{return}\;\Phi,\;\Phi^{-1},\;\left[\mathcal{K}_d, \dots , \mathrm{B}_{m}
\right],\;V_\theta(\boldsymbol{z}) $
\State $\textbf{end function}$
\State $\textbf{function}\;\;\texttt{FALSIFICATION}(X)$
\State $\quad\text{Impose conditions defined in Eqn. \eqref{eqn:falsification_constraint} }
$
\State $\quad\text{Use SMT solver}\; \text{to verify the conditions}$
\State $\quad\textbf{return}\;\text{satisfiability} $
\State $\textbf{end function}$
\State $\textbf{function}\;\;\texttt{CONTROL}(\boldsymbol{x}_0)$\Comment{Stabilizing feedback controller}
\State $\quad \Phi,\;\Phi^{-1},\;\left[\mathcal{K}_d, \dots , \mathrm{B}_{m}
\right],\;V_\theta(\boldsymbol{z})\gets\texttt{LEARNING}(X)$
\State $\quad\textbf{Repeat:}$
\State $\quad \boldsymbol{u}\gets\text{Eqn. }\eqref{eqn:sontag_formula},\quad U\gets \textnormal{Append}(\boldsymbol{u})$\Comment{Universal Sontag's formula}
\State $\quad \boldsymbol{x}_{\text{next}}\gets \text{Apply feedback control law}\;\boldsymbol{u}\;\text{to discrete system (Eqn.  }\eqref{eqn:discretize_of_nonlinear_system})$\Comment{Propagate the state}
\State $\quad \boldsymbol{x}\gets\boldsymbol{x}_{\text{next}}$
\State $\quad\textbf{until}\; \text{convergence}$
\State $\quad\textbf{return}\;U $
\State $\textbf{end function}$
\State $\textbf{function}\;\;\text{\texttt{MAIN}()}$
\While{\text{Satisfiable}}
\State $\quad\text{Add counterexamples to}\; X$
\State $\quad \Phi,\;\Phi^{-1},\;\left[\mathcal{K}_d, \dots , \mathrm{B}_{m}
\right],\;V_\theta(\boldsymbol{z})\gets\texttt{LEARNING}(X)$
\State $\quad S\gets\texttt{FALSIFICATION}(X)$
\EndWhile
\State $U\gets\texttt{CONTROL}(\boldsymbol{x}_0)$
\State $\textbf{end function}$
 \end{algorithmic}
\label{alg:learn_bilinear}
\end{algorithm}
\section{Results\label{sec:results}}
In this section, we present the experimental results of our proposed approach on various nonlinear control problems. The learning framework is implemented in PyTorch for the simulations and our implementation follows Algorithm \ref{alg:learn_bilinear}. The dReal package (SMT solver) is used to generate counterexamples for learning a valid CLF. The code for the numerical experiments can be found at \url{https://github.com/Vrushabh27/Neural-Koopman-Lyapunov-Control}. In contrast to the approach proposed in \cite{chang2020neural,mehrjou2020neural_lyapunov_redesign,ravanbakhsh2019learning_lyapunov_from_demonstrations,jin2020neural_control_policy,abate2020lyapunov_formal_synthesis,dai2021lyapunov_stable_nn_control}, the main advantage of using our proposed method is two-fold. First, Koopman operator theory allows us to analyze a unknown {control-affine} nonlinear system via a learned higher dimensional Koopman based bilinear system. Second, unlike recent methods which restrict themselves to linear feedback controllers or learn a neural network based controller, our approach provides provable guarantee for the existence of a stabilizing controller for the unknown {control-affine} nonlinear system. 

The training was performed for at least 900 epochs with Adam optimizer before the falsifier was used to generate counterexamples (that violate CLF conditions) which were then added to the training data. The activation function used for the encoder, decoder and the CLF is $\texttt{Tanh}$ with learning rate set to $10^{-3}$ and the Mean Squared Error (MSE) loss for all the examples. The analytical expressions for $V_\theta(\boldsymbol{z})$ and $\Phi(\boldsymbol{x})$ are required to generate counterexamples by a SMT verifier (falsifier \eqref{eqn:falsification_constraint}) and compute the feedback stabilizing controller (Eqn. \eqref{eqn:sontag_formula}). The expressions for $V_\theta(\boldsymbol{z})$ and $\Phi(\boldsymbol{x})$ are represented by the following recursive relations:
\begin{align}
& \textbf{CLF:}\quad V_\theta(\boldsymbol{z})=\texttt{tanh}(W^v_{h^v+1}y_{i+1}+b^v_{h+1})\quad\text{where}\quad y_{i}=\texttt{tanh}(W^v_iy_{i-1}+b^v_i),\;\;i=h^v,\;\;y^v_1=\boldsymbol{z}\;\;\nonumber\\
&\textbf{Encoder:}\quad\Phi(\boldsymbol{x})=\texttt{tanh}(W^e_{h^e+1}y_{i+1}+b^e_{h^e+1})\quad\text{where}\quad y_{i}=\texttt{tanh}(W^e_iy_{i-1}+b^e_i),\;\;i=h^e,\;\;y^e_1=\boldsymbol{x}\nonumber
\end{align}
where $h^v,\;W^v_i,\;b^v_i$ and $h^e,\;W^e_i,\;b^e_i$ denote the number of hidden layers, weights and biases of the CLF and encoder respectively. In order to train the encoder, decoder and the CLF, we assume that a black-box simulator of the unknown nonlinear dynamics \eqref{eqn:discretize_of_nonlinear_system} is available where $\boldsymbol{x}_{k+1}$ is returned given $\boldsymbol{x}_k$ and $\boldsymbol{u}_k$.

For the pendulum system and the Van Der Pol oscillator, we use the encoder and CLF with 1 hidden layer of 6 units each  whereas for the cart pole system and the spacecraft rendezvous problem, we use the encoder and CLF with 2 hidden layers of 32 units each. For the decoder, we use a neural network with 2 hidden layers of 16 units each for all experiments.

\textbf{Hyperparameter tuning}; The hyperparameters $\alpha_1,\;\alpha_2,\;\alpha_3$ and $\alpha_4$ were tuned based on the combination of the controller performance and the empirical loss on the training data. We observed that for the pendulum and Van Der Pol system, $\alpha_1=\alpha_3=0.001$, $\alpha_2=2$ and $\alpha_4=1$ yielded best results whereas for the cart pole and spacecraft rendezvous system we obtained the best results for $\alpha_1=\alpha_3=0.05$, $\alpha_2=3$ and $\alpha_4=1$. 

In the following subsections, we consider two popular academic nonlinear systems followed by two real world nonlinear systems.
\subsection{Pendulum system}
\begin{figure}[H]
 \captionsetup[subfigure]{justification=centering}
 \centering
 \begin{subfigure}{0.47\textwidth}
{\includegraphics[width=6cm]{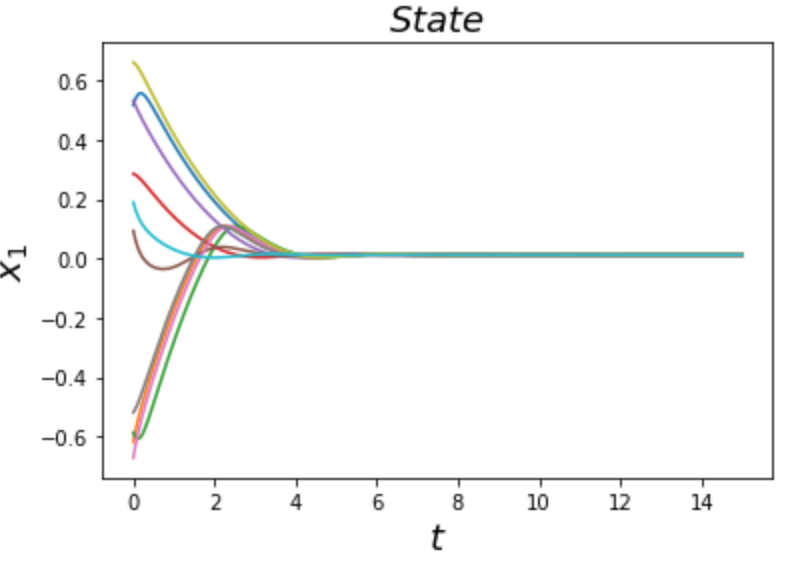}}
\caption{$x_1$}
\label{fig:pendulum_states_1}
 \end{subfigure}
 \begin{subfigure}{0.45\textwidth}
 \includegraphics[width=6cm]{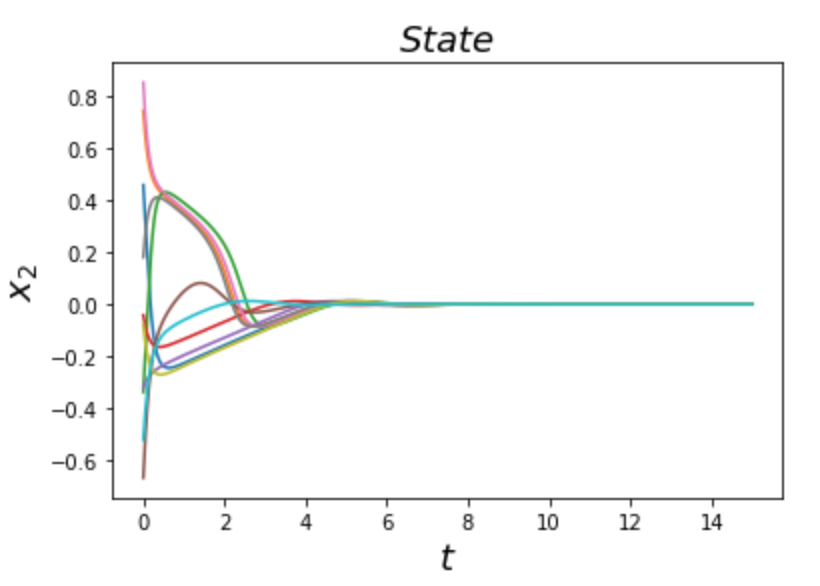}
 \caption{$x_2$}
\label{fig:pendulum_states_2}
 \end{subfigure}
 \begin{subfigure}{0.47\textwidth}
 \includegraphics[width=6cm]{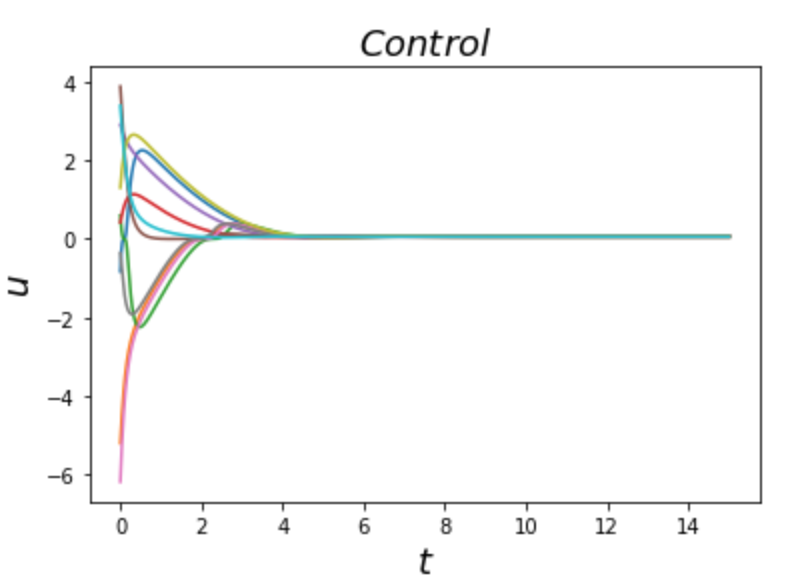}
\caption{$u$}
\label{fig:pendulum_u}
\end{subfigure}
\begin{subfigure}{0.44\textwidth}
 \includegraphics[width=6cm]{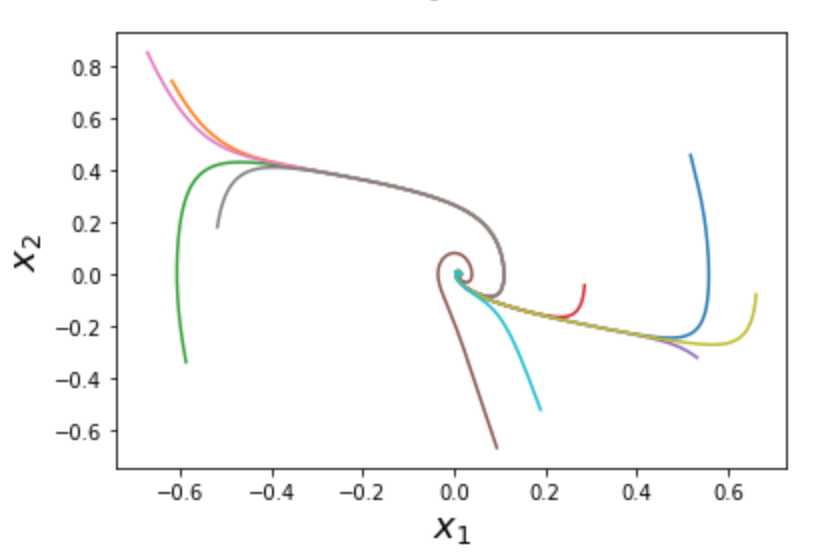}
\caption{}
\label{fig:phase_space_pendulum}
\end{subfigure}
 \caption{The evolution of the state and control input with time for the pendulum system}
\label{fig:pendulum}
\end{figure}

The pendulum dynamics is given by
\begin{align}
    &\dot{{x}}_1={x}_2,\quad\quad\dot{{x}}_2=-\frac{mg}{l}\text{sin}({x}_1)+u
\end{align}
where $\boldsymbol{x}=[x_1,\;x_2]^\mathrm{T}$ is the state, $u$ is the control input, $m$ and $l$ is the mass and length of the pendulum respectively. We set the sampling time $T$ to be $0.005s$. Note that the origin corresponds to an equilibrium of the pendulum system in the control-free case. The state domain $\mathcal{X}=\{\boldsymbol{x}:\boldsymbol{x}_{\text{lb}}\leq\boldsymbol{x}\leq\boldsymbol{x}_{\text{ub}}\}$ where $\boldsymbol{x}_{\text{lb}}=[-1,\;-1]^\mathrm{T}$ and $\boldsymbol{x}_{\text{ub}}=[1,\;1]^\mathrm{T}$. Figs. \ref{fig:pendulum_states_1} and \ref{fig:pendulum_states_2} shows the evolution of the ten trajectories whose initial conditions are randomly sampled from $\mathcal{X}$ and Fig. \ref{fig:pendulum_u} shows the evolution of the stabilizing control inputs (based on the learned CLF, Koopman observables $\boldsymbol{z}$ and Sontag's formula \eqref{eqn:sontag_formula}) which were applied to these trajectories. Fig. \ref{fig:phase_space_pendulum} shows the trajectories in the phase space.

\subsection{Van Der Pol oscillator}
\begin{figure}[H]
 \captionsetup[subfigure]{justification=centering}
 \centering
 \begin{subfigure}{0.47\textwidth}
{\includegraphics[width=6cm]{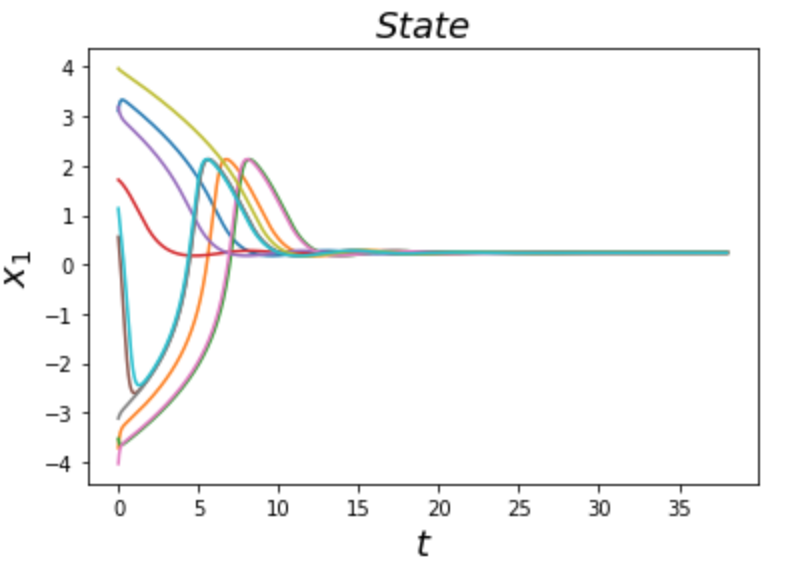}}
\caption{$x_1$}
\label{fig:x1_vanderpol}
 \end{subfigure}
 \begin{subfigure}{0.45\textwidth}
 \includegraphics[width=6cm]{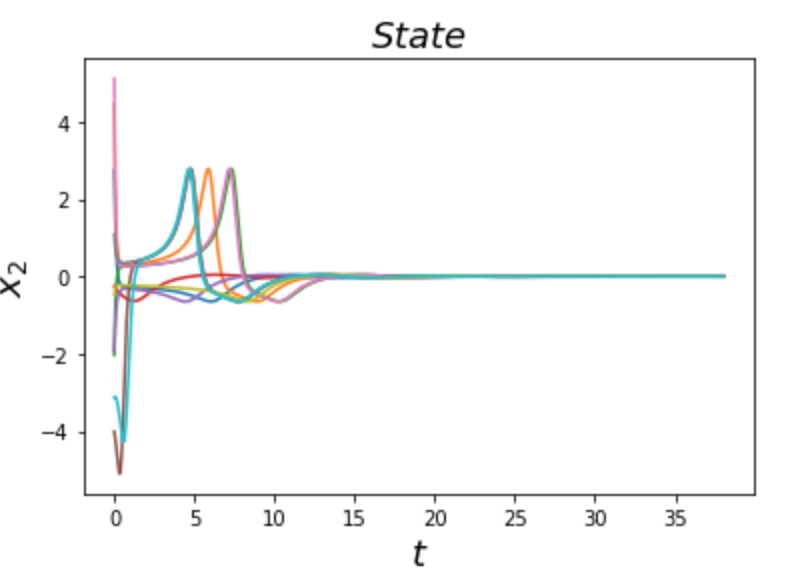}
 \caption{$x_2$}
\label{fig:x2_vanderpol}
 \end{subfigure}
 \begin{subfigure}{0.47\textwidth}
 \includegraphics[width=6cm]{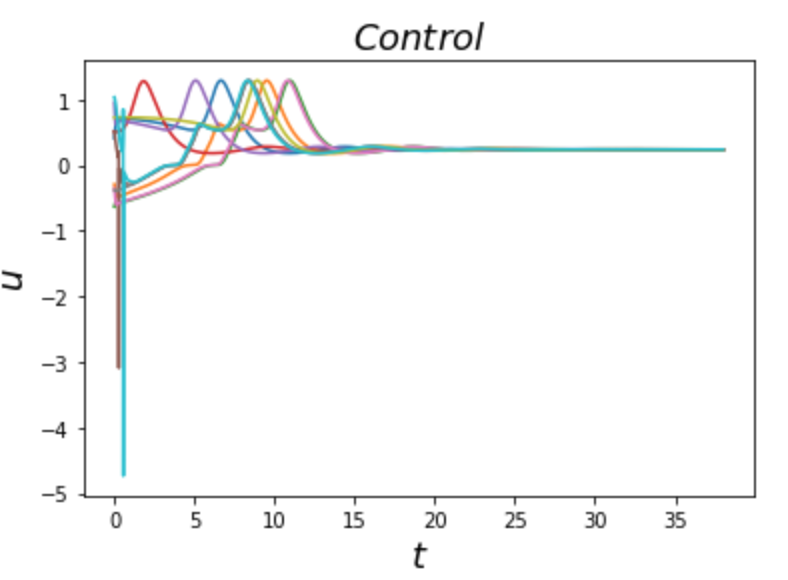}
\caption{$u$}
\label{fig:u_vanderpol}
\end{subfigure}
\begin{subfigure}{0.44\textwidth}
 \includegraphics[width=6cm]{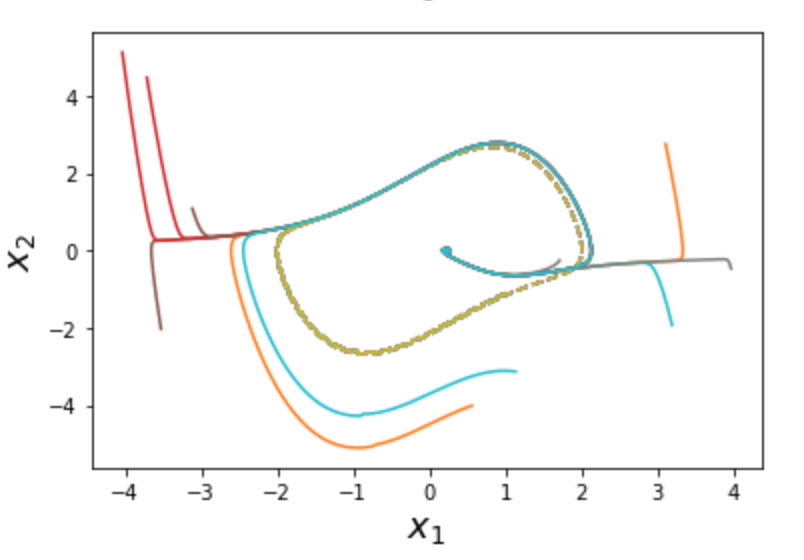}
\caption{}
\label{fig:phase_space_vander_pol}
\end{subfigure}
 \caption{The evolution of the state and control input with time for the Van Der Pol oscillator}
\label{fig:van_der_pol}
\end{figure}

Next, we consider the Van Der Pol oscillator system whose governing equations are given by
\begin{align}
    &\dot{{x}}_1={x}_2,\quad\quad\dot{{x}}_2=(1-{x}^2_1){x}_2+{x}_1+u
\end{align}
where $\boldsymbol{x}=[x_1,\;x_2]^\mathrm{T}\in\mathbb{R}^2$ is the state and $u\in\mathbb{R}$ is the control input for the Van Der Pol oscillator. We define the state domain $\mathcal{X}=\{\boldsymbol{x}:\boldsymbol{x}_{\text{lb}}\leq\boldsymbol{x}\leq\boldsymbol{x}_{\text{ub}}\}$ (the inequality denotes the element wise comparison) where $\boldsymbol{x}_{\text{lb}}=[-10,\;-10]^\mathrm{T}$ and $\boldsymbol{x}_{\text{ub}}=[10,\;10]^\mathrm{T}$. We set the sampling time $T=0.01$. Figs. \ref{fig:x1_vanderpol} and \ref{fig:x2_vanderpol} show the evolution of the states of the Van Der Pol oscillator with initial conditions $\boldsymbol{x}_0$ sampled from the set $\mathcal{X}$. Once a valid CLF is learned from training the neural network, it is used in the Sontag's formula to design a stabilizing feedback controller that would steer the state from $\boldsymbol{x}_0$ to origin.

The control inputs are shown in Fig. \ref{fig:u_vanderpol} Note that the origin is the unstable equilibrium. If no control input is applied, the trajectories will converge to a limit cycle (shown by green dotted curve in Fig. \ref{fig:phase_space_vander_pol}), irrespective of the initial state.
\subsection{Cart pole system}
\begin{figure}[H]
 \captionsetup[subfigure]{justification=centering}
 \centering
 \begin{subfigure}{0.47\textwidth}
{\includegraphics[width=6cm]{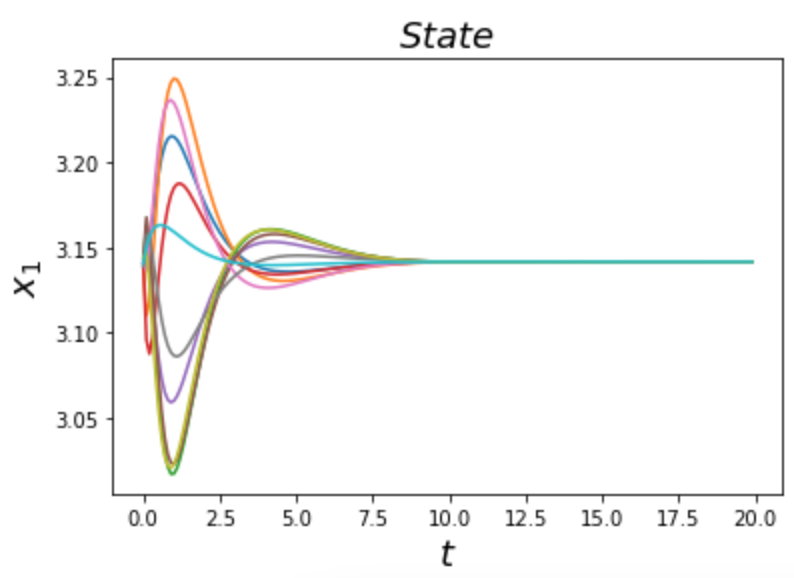}}
\caption{$x_1$}
\label{fig:x1_cart}
 \end{subfigure}
 \begin{subfigure}{0.45\textwidth}
 \includegraphics[width=6cm]{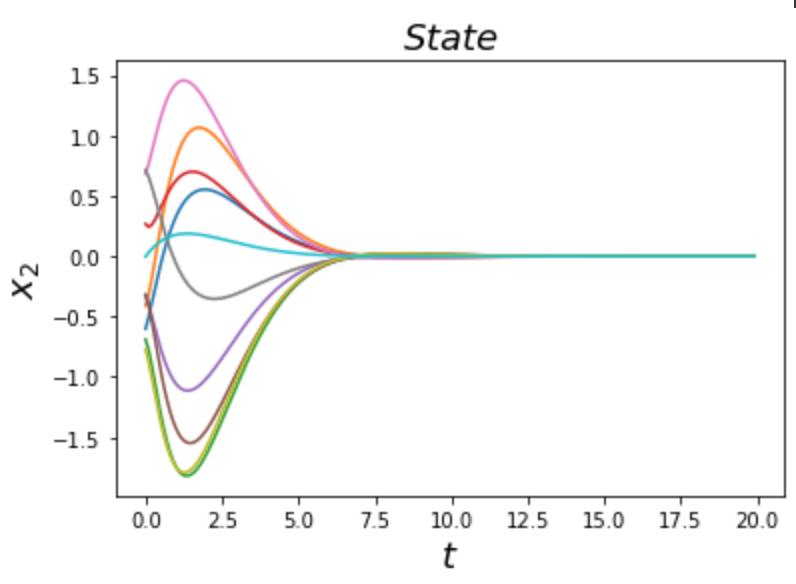}
 \caption{$x_2$}
\label{fig:x2_cart}
 \end{subfigure}
  \begin{subfigure}{0.45\textwidth}
 \includegraphics[width=6cm]{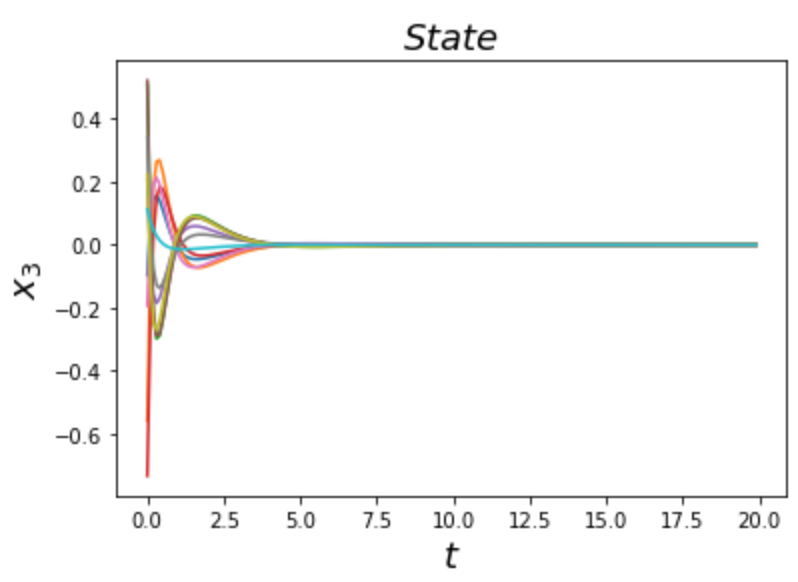}
 \caption{$x_3$}
\label{fig:x3_cart}
 \end{subfigure}
  \begin{subfigure}{0.45\textwidth}
 \includegraphics[width=6cm]{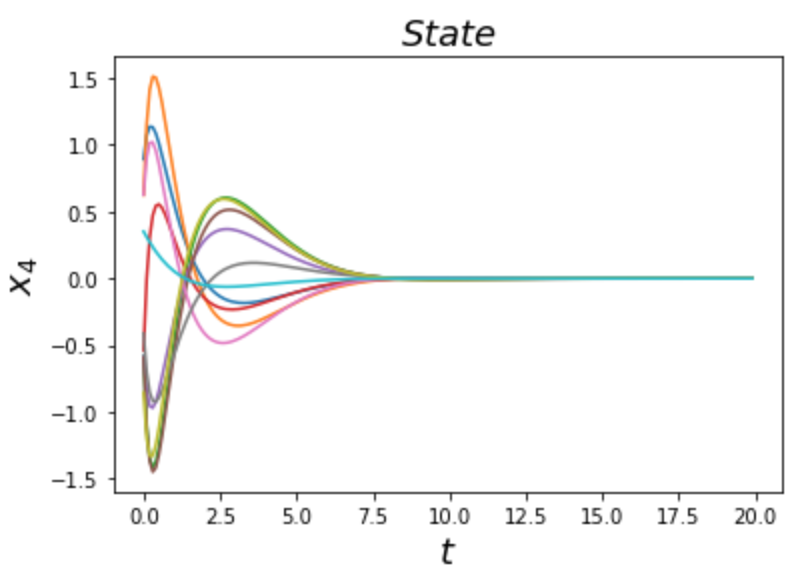}
 \caption{$x_4$}
\label{fig:x4_cart}
 \end{subfigure}
 \begin{subfigure}{0.47\textwidth}
 \includegraphics[width=6cm]{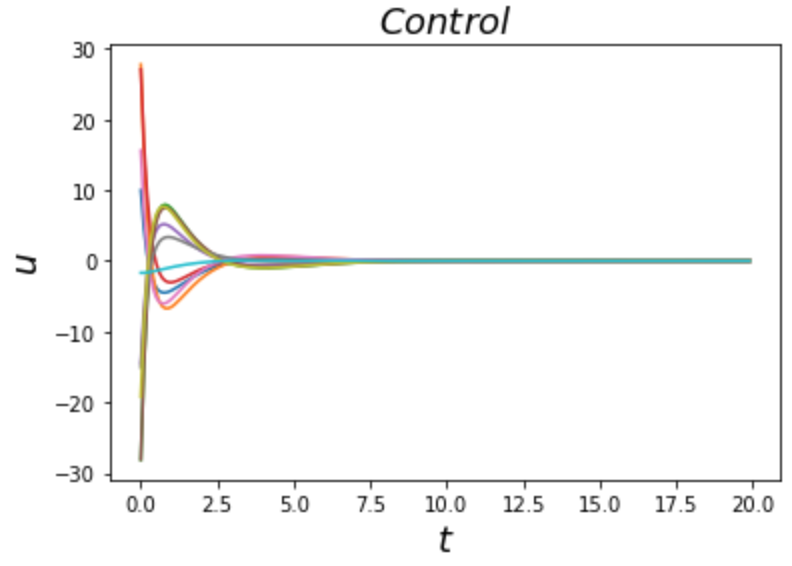}
\caption{$u$}
\label{fig:u_cart}
\end{subfigure}
 \caption{The evolution of the state and control input with time for the cart pole system}
\label{fig:cart_pole}
\end{figure}

The cart pole system is a fully underactuated system with one control input and two degrees of freedom (DOF). Due to its highly nonlinear structure, it is used to validate nonlinear controllers. Cart pole systems can find many applications such as rocket propellers, self balancing robots, stabilization of ships etc. The cart pole dynamics is given as follows:
\begin{align}
\dot{x} =\frac{u+m_{p} \sin \theta\left(l \dot{\theta}^{2}-g \cos \theta\right)}{m_{c}+m_{p}(\sin \theta)^{2}},\quad\quad
\ddot{\theta} =\frac{u \cos \theta+m_{p} l \dot{\theta}^{2} \cos \theta \sin \theta-\left(m_{c}+m_{p}\right) g \sin \theta}{l\left(m_{c}+m_{p}(\sin \theta)^{2}\right)}
\end{align}
where $m_c=4$ is the mass of the cart, $l=1$ and $m_p=1$ are the length and mass of the pole respectively, $\boldsymbol{x}=[x_1,\;x_2,\;x_3,\;x_4]^\mathrm{T}=[x,\;\theta,\;\dot{x},\;\dot{\theta}]^\mathrm{T}$ and ${u}$ is the control input which controls the linear velocity of the cart. The objective is to steer the initial state of the cart to an upright position. We set $T=0.005$. Fig. \eqref{fig:cart_pole} shows the convergence of the trajectories starting from a set of ten randomly selected initial conditions to the origin when a stabilizing feedback controller is applied to the unknown {control-affine} nonlinear system. 

\subsection{Spacecraft rendezvous operation}
\begin{figure}[H]
 \captionsetup[subfigure]{justification=centering}
 \centering
 \begin{subfigure}{0.47\textwidth}
{\includegraphics[width=6cm]{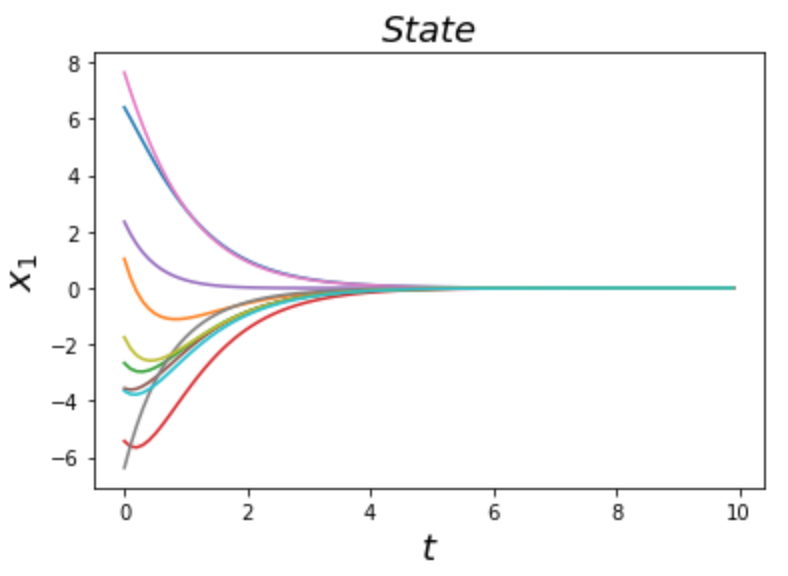}}
\caption{$x_1$}
\label{fig:x1_sat}
 \end{subfigure}
 \begin{subfigure}{0.45\textwidth}
 \includegraphics[width=6cm]{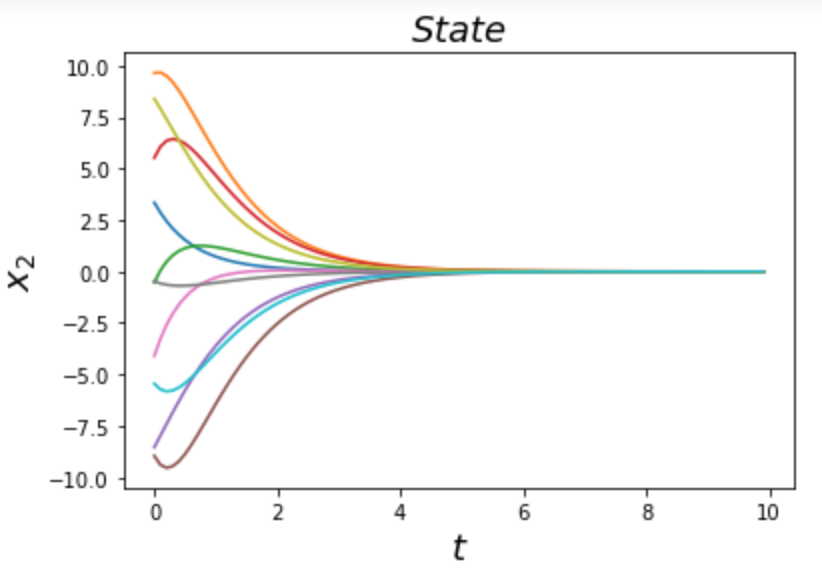}
 \caption{$x_2$}
\label{fig:x2_sat}
 \end{subfigure}
  \begin{subfigure}{0.45\textwidth}
 \includegraphics[width=6cm]{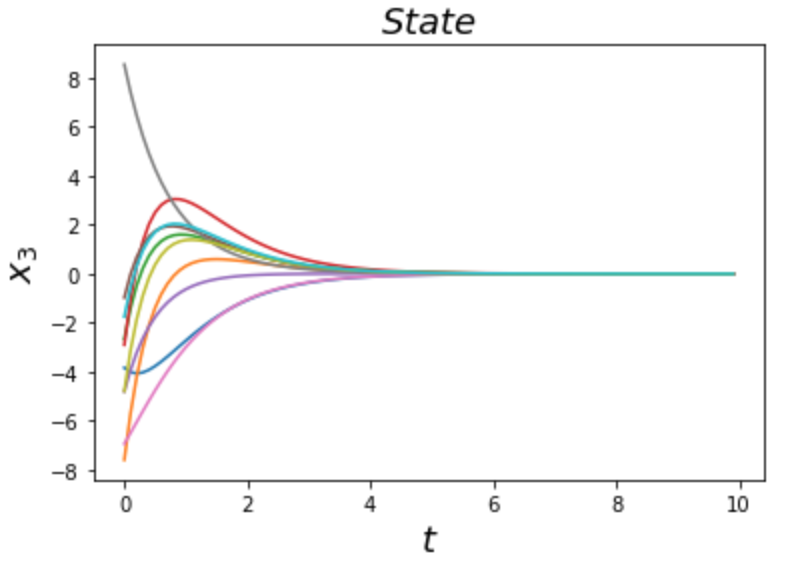}
 \caption{$x_3$}
\label{fig:x3_sat}
 \end{subfigure}
  \begin{subfigure}{0.45\textwidth}
 \includegraphics[width=6cm]{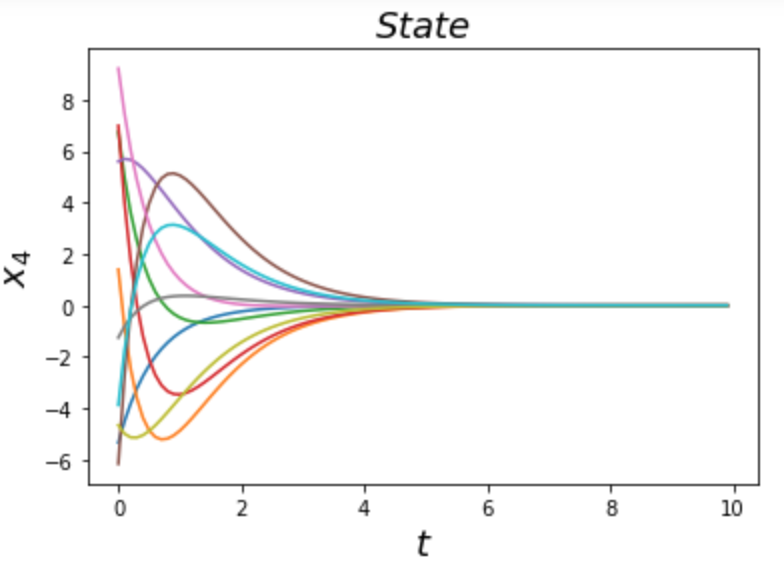}
 \caption{$x_4$}
\label{fig:x4_sat}
 \end{subfigure}
 \begin{subfigure}{0.47\textwidth}
 \includegraphics[width=6cm]{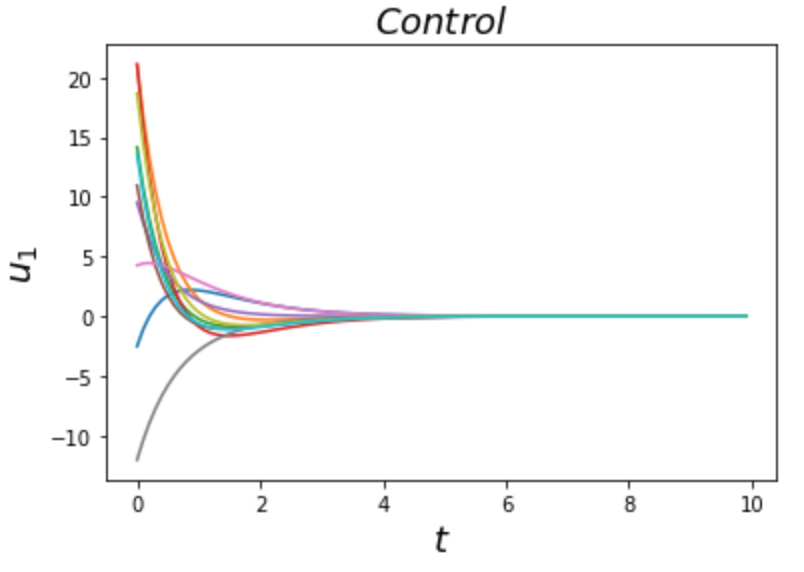}
\caption{$u_1$}
\label{fig:u1_sat}
\end{subfigure}
 \begin{subfigure}{0.47\textwidth}
 \includegraphics[width=6cm]{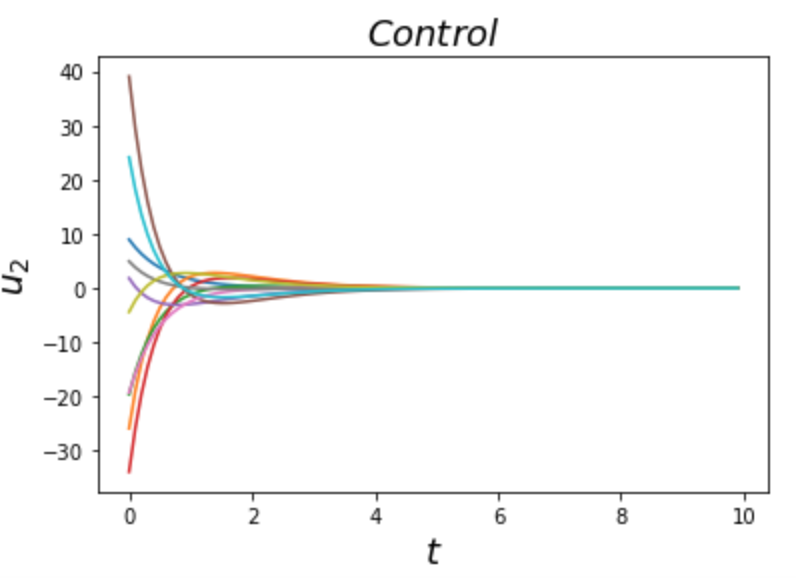}
\caption{$u_2$}
\label{fig:u2_sat}
\end{subfigure}
 \caption{The evolution of the state and control input with time for the spacecraft rendezvous operation}
\label{fig:rendezvous}
\end{figure}

Last, we consider the spacecraft rendezvous operation whose governing equations are given by the Hill Clohessy Wiltshire (HCW) equations as follows:
\begin{align}
\ddot{x} =3 n^{2} x+2 n \dot{y}+u_1,\quad\quad
\ddot{y} &=-2 n \dot{x}+u_2
\end{align}
where $\boldsymbol{x}=[x,\;y,\;\dot{x},\;\dot{y}]^\mathrm{T}$, $\boldsymbol{u}=[u_1,\;u_2]^\mathrm{T}$,  $n=\sqrt{\mu/a^3}$, $a=6793137$ (low earth orbit) is the length of the semi-major axis, $\mu=3.986\times 10^{14}$ is the gravitational constant. For more details on the HCW model, the reader may refer to \cite{curtis2013orbital}. We set $T=0.005$. Figs. \ref{fig:x1_sat}-\ref{fig:x4_sat} shows the evolution of the closed-loop trajectories starting from initial conditions which are sampled uniformly from $[-10,10]^4$. Figs. \ref{fig:u1_sat} and \ref{fig:u2_sat} shows the evolution of stabilizing feedback control inputs for the rendezvous problem.

\begin{remark}
\normalfont Note that depending on the complexity of the system dynamics, the dimension of the lifted space $N$ can be increased or decreased. Further, it was observed that with increase in $N$, the computational time required for SMT solver to find counterexamples for computing the CLF increases. 
\end{remark}
\section{Conclusion\label{sec:conclusion}}
We proposed a Koopman operator based learning framework to compute a lifted bilinear system that serves as a higher dimensional representation of an unknown {control-affine} nonlinear system and design a stabilizing feedback controller based on a Control Lyapunov Function (CLF) which is computed under the same learning framework. Our approach simultaneously learns 1) the matrices that determine the state space model representation of the bilinear system, 2) the Koopman based observables and 3) a valid CLF by using a learner and a falsifier. The learned CLF is then used to design a stabilizing feedback controller (based on the learned Koopman bilinear system) which is then applied to the control-affine nonlinear system. Numerical experiments are provided to validate that our proposed class of learning-based stabilizing feedback controllers is able to stabilize the unknown control-affine nonlinear system. A particularly exciting direction for our future work is to use the learned bilinear model to design robust control laws for the unknown nonlinear system which can account for disturbances acted upon the latter system as well as modelling errors and uncertainties. Another possible direction is to extend the results presented herein to stochastic nonlinear control systems. 

\section*{Acknowledgments}
This research has been supported in part by NSF award CMMI-1937957.

\bibliographystyle{unsrt}  
\bibliography{main}

\end{document}